\documentclass[aps,prb,amsmath,superscriptaddress,twocolumn]{revtex4-2}

\usepackage{color}
\usepackage{graphicx}
\usepackage{epsfig,psfrag}
\usepackage{amsmath}
\usepackage{amssymb}
\usepackage{amsbsy}
\usepackage{amsthm}
\usepackage{amsfonts}
\usepackage{wasysym}
\usepackage{bbm}
\usepackage{tabularx}
\usepackage{euscript}
\usepackage{enumerate}
\usepackage{amsfonts}
\usepackage{exscale}
\usepackage{bbold}
\usepackage{float}
\usepackage{slashed}
\usepackage{subfigure}
\usepackage{comment}
\usepackage[colorlinks,citecolor=blue]{hyperref}
\usepackage{soul}
\usepackage{ulem}

\newcommand{\bea}{\begin{eqnarray}}
\newcommand{\eea}{\end{eqnarray}}
\newcommand{\ba}{\begin{array}}
\newcommand{\ea}{\end{array}}

\def\bea{\begin{eqnarray}}
\def\eea{\end{eqnarray}}

\def\ii{\mathfrak{i}}

\AtBeginDocument{%
    \newwrite\bibnotes
    \def\bibnotesext{Notes.bib}
    \immediate\openout\bibnotes=\jobname\bibnotesext
    \immediate\write\bibnotes{@CONTROL{REVTEX41Control}}
    \immediate\write\bibnotes{@CONTROL{%
    apsrev41Control,author="08",editor="1",pages="1",title="0",year="1"}}
     \if@filesw
     \immediate\write\@auxout{\string\citation{apsrev41Control}}%
    \fi
}%

\begin{document}

\title{Yang-Lee edge singularity triggered entanglement transition}

\author{Shao-Kai Jian}\thanks{These authors contributed equally to this work.}
\affiliation{Condensed Matter Theory Center, Department of Physics, University of Maryland, College Park, Maryland 20742, USA}

\author{Zhi-Cheng Yang}\thanks{These authors contributed equally to this work.}
\affiliation{Joint Quantum Institute, University of Maryland, College Park, Maryland 20742, USA}
\affiliation{Joint Center for Quantum Information and Computer Science, University of Maryland, College Park, Maryland 20742, USA}

\author{Zhen Bi}
\affiliation{Department of Physics, The Pennsylvania State University, University Park, Pennsylvania 16802, USA}

\author{Xiao Chen}
\affiliation{Department of Physics, Boston College, Chestnut Hill, MA 02467, USA}

\begin{abstract}
We show that a class of $\mathcal{PT}$ symmetric non-Hermitian Hamiltonians realizing the Yang-Lee edge singularity exhibits an entanglement transition in the long-time steady state evolved under the Hamiltonian. Such a transition is induced by a level crossing triggered by the critical point associated with the Yang-Lee singularity
and hence is first-order in nature. At the transition, the entanglement entropy of the steady state jumps discontinuously from a volume-law to an area-law scaling. We exemplify this mechanism using a one-dimensional transverse field Ising model with additional imaginary fields, as well as the spin-1 Blume-Capel model and the three-state Potts model. We further make a connection to the forced-measurement induced entanglement transition in a Floquet non-unitary circuit subject to continuous measurements followed by post-selections. Our results demonstrate a new mechanism for entanglement transitions in non-Hermitian systems harboring a critical point.

\end{abstract}
\date{\today}

\maketitle

{\it Introduction}.-- The dynamics of entanglement provides a quantum information perspective on the non-equilibrium dynamics of many-body systems. For chaotic systems -- Hamiltonians or random unitary circuits -- the entanglement entropy under time evolution typically saturates to a volume-law scaling with the subsystem size, indicating thermalization at late times~\cite{PhysRevLett.111.127205, PhysRevE.91.062128, PhysRevB.95.094302, PhysRevX.7.031016, PhysRevB.99.174205}. However, this scenario is altered once the system is coupled to the environment and one tracks an individual quantum trajectory at a time. A minimally structured setup capturing the latter scenario consists of a random unitary circuit interspersed with weak measurements, and a particular sequence of measurement outcome is recorded. Remarkably, such hybrid random unitary circuits feature an entanglement phase transition from a volume-law phase to an area-law phase, as the measurement rate is varied~\cite{PhysRevB.98.205136, PhysRevB.100.134306, PhysRevB.99.224307, PhysRevX.9.031009, PhysRevX.10.041020, PhysRevLett.125.070606, fan2020self, PhysRevLett.125.030505, iaconis2020measurement, sang2020measurement,lavasani2020measurement, lavasani2020topological,ippoliti2020entanglement}. In (1+1) dimension, this entanglement transition in hybrid random unitary circuits is generically a continuous one exhibiting similar properties vis-\`{a}-vis certain non-unitary conformal field theory (CFT) upon mapping to a statistical-mechanics model~\cite{PhysRevB.100.134306, PhysRevB.100.134203, PhysRevB.101.104302, PhysRevX.9.031009, fan2020self, li2020conformal, PhysRevB.101.104301}.

One expects that the existence of temporal randomness (randomness in gate compositions, measurement locations, and measurement outcomes) is crucial for the universality class of hybrid random unitary circuits, as randomness is typically relevant in lower dimensions~\cite{harris1974effect}. It is thus of great interest to ask whether there can be new possibilities or mechanisms for entanglement transitions in systems where all randomness is removed. One such example is a system subject to continuous weak measurements, and one post-selects a trajectory with a specified outcome. The time evolution in this case can be generated by a non-Hermitian non-random Hamiltonian with an imaginary field, for which one may infer the long-time steady state~\cite{footnote2} solely from the eigenvalues and eigenstates. Recent works have identified entanglement transitions in non-Hermitian systems of free fermions~\cite{biella2020many}, chaotic spin chains~\cite{gopalakrishnan2020entanglement}, and the Sachdev-Ye-Kitaev chain in the large $N$ limit~\cite{liu2020non}. Nonetheless, the existence and nature of entanglement transitions in non-Hermitian systems remain to be better understood.

\begin{figure}[t]
	\centering
	\subfigure[]{\label{fig:spectrum_transition} 
		\includegraphics[width=0.47\textwidth]{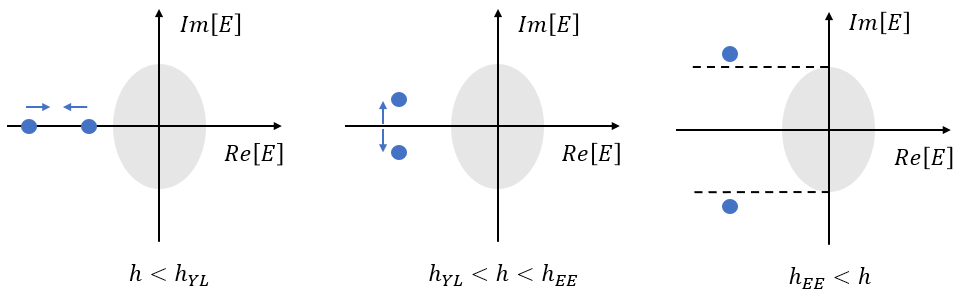}} \\
	\subfigure[]{\label{fig:transition_imag0}
		\includegraphics[width=0.22\textwidth]{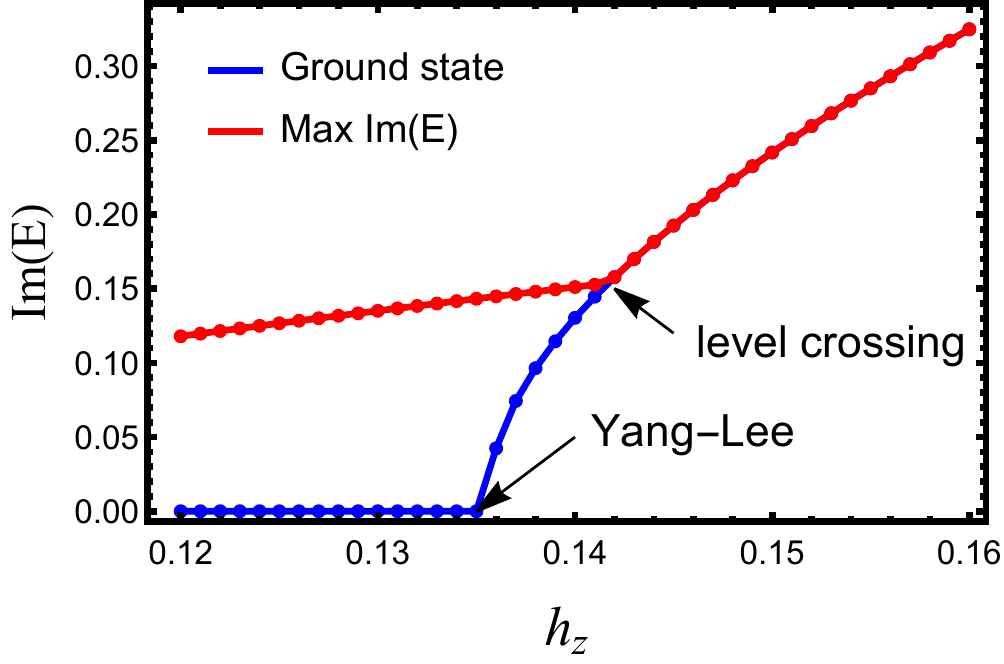}} \quad
	\subfigure[]{ \label{fig:transition_order0}
		\includegraphics[width=0.215\textwidth]{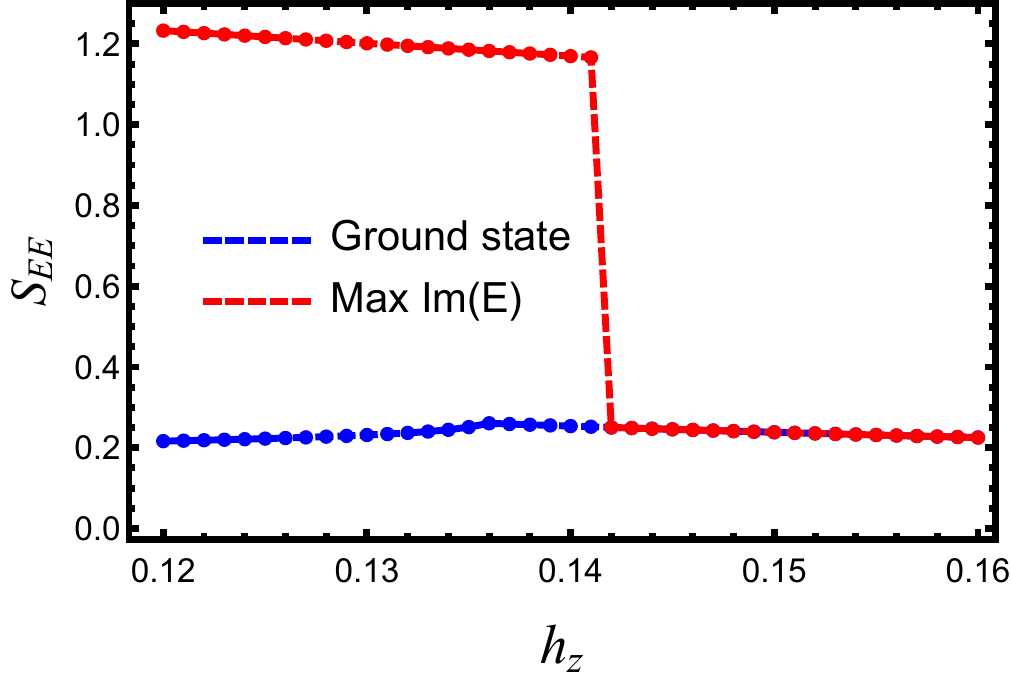}} 
	\caption{(a) A schematic of the coalescing-splitting process between the ground state (state with the smallest real eigenvalue) and the first excited state (blue dots). The grey region denotes the (complex) eigenvalues of the rest of the eigenstates.
		(b)\&(c) The imaginary part of the eigenenergy (b), and the half-chain entanglement entropy (c) of the ground state (blue) and state with the maximal ${\rm Im}(E)$ (red) as a function of $h_z$. We choose $J = 0.4$, $J_2=0.1$, $\Gamma = 1, h_y = 0$, and system size $L=10$
		with periodic boundary condition.}
	\label{fig:2}
\end{figure}

In this work, we demonstrate a new mechanism leading to a first-order entanglement transition in a class of $\mathcal{PT}$ symmetric non-Hermitian Hamiltonians, whose ground state (hereafter referring to the state with the smallest real eigenvalue) undergoes a continuous phase transition belonging to the Yang-Lee universality class~\cite{PhysRev.87.404, PhysRev.87.410, Fisher:1978yang, Cardy:1985conformal, matsumoto2020embedding}. 
The Hamiltonian we consider takes the form
\begin{align}
    H=H_1+i H_2
\end{align}
where $H_1$ is a Hermitian interacting Hamiltonian and $H_2$ denotes an imaginary field. If the ground state of $H_1$ is in the paramagnetic phase, due to $\mathcal{PT}$ symmetry, as the imaginary field increases,
the ground state and first excited state energies remain real until the gap closes at the Yang-Lee critical point, after which they start splitting in pairs along the imaginary axis. 
The development of magnetic ordering in the ground state past the critical point continues driving the growth of its imaginary eigenenergy, which eventually leads to a level crossing along the imaginary axis with some (typically) highly excited state. See Fig.~\ref{fig:spectrum_transition} for an illustration~\cite{eignenergy}. 
Since the long-time steady state under time evolution is governed by the right eigenstate with the largest imaginary eigenenergy, this level crossing signals a discontinuous jump in the steady state entanglement from a volume-law to an area-law scaling, because the ground state of $H_1$ has an area-law scaling entanglement whereas a typical excited state of $H_1$ has a volume-law scaling entanglement. 
In contrast, if the ground state of $H_1$ is in the ordered phase, the steady state is area-law entangled immediately as we introduce $H_2$, and hence there is no entanglement transition. 
We exemplify the above scenario using various one dimensional quantum spin models with additional imaginary fields, and further make a connection to the forced-measurement induced entanglement transition.

{\it Yang-Lee edge singularity}.-- The ferromagnetic phase transition of the classical Ising model in an external magnetic field $h$ can be understood from the zeros of the partition function on the complex $h$ plane. Above the critical temperature $T>T_c$, all zeros are distributed along the imaginary axis $|{\rm Im}(h)| \geq h_{YL}(T)$, with $h_{YL}(T)$ vanishing as $T$ approaches $T_c$~\cite{PhysRev.87.404, PhysRev.87.410}. The Yang-Lee edge singularity $h_{YL}$ in fact can be regarded as a conventional critical point described by a $\phi^3$ field theory with imaginary couplings~\cite{Fisher:1978yang, PhysRevLett.27.1439} (see Supplemental Materials for a brief review~\cite{SM}). As such, the Yang-Lee singularity can be alternatively realized as a \textit{quantum} phase transition in a $1+1d$ non-Hermitian quantum Hamiltonian~\cite{von:1991critical}, where the energy gap between the two states with lowest real eigenenergies closes on the real axis and reopens on the imaginary axis in a universal manner across the transition (see Fig.~\ref{fig:2}(a)). 
By tracking the evolution of the eigenenergy levels on the complex plane across the Yang-Lee critical point, we will show that there must be a first-order entanglement transition induced by a level-crossing along the imaginary axis of the eigenenergy spectrum.

Consider a one-dimensional (1D) transverse field Ising model with next-nearest-neighbor couplings and in the presence of imaginary fields
\begin{equation}
\label{eq:hamiltonian}
	H =
   - \sum_{i=1}^L (J \sigma^z_i \sigma^z_{i+1}+J_2 \sigma^z_i \sigma^z_{i+2} + \Gamma \sigma^x_i + i h_z \sigma^z_i + i h_y \sigma^y_i ),
\end{equation}
where $\sigma^{x,y,z}$ denotes Pauli matrices, and $J, J_2, \Gamma,  h_z,  h_y>0$ are real parameters. We have included a $J_2$ term such that Hamiltonian~(\ref{eq:hamiltonian}) is non-integrable in the absence of imaginary fields; nevertheless, both the Yang-Lee singularity and the entanglement transition persist when $J_2=0$.
This model, despite non-Hermitian, has a generalized $\mathcal{PT}$ symmetry which we define below.
Therefore, the eigenvalues of Hamiltonian~(\ref{eq:hamiltonian}) must either be real, or come in complex conjugate pairs. A single eigenenergy cannot leave the real axis without coalescing with a partner and then splitting in pairs. In particular, this is also the case for the ground state of Hamiltonian~(\ref{eq:hamiltonian}). In spite of the similarity between Hamiltonian~(\ref{eq:hamiltonian}) and the models studied in Refs.~\cite{biella2020many, gopalakrishnan2020entanglement}, we point out that those models do \textit{not} have a Yang-Lee singularity, and hence the mechanisms for the entanglement transition therein are completely different from Hamiltonian~(\ref{eq:hamiltonian}).

\begin{figure}[t]
	\centering
	\includegraphics[width=0.3\textwidth]{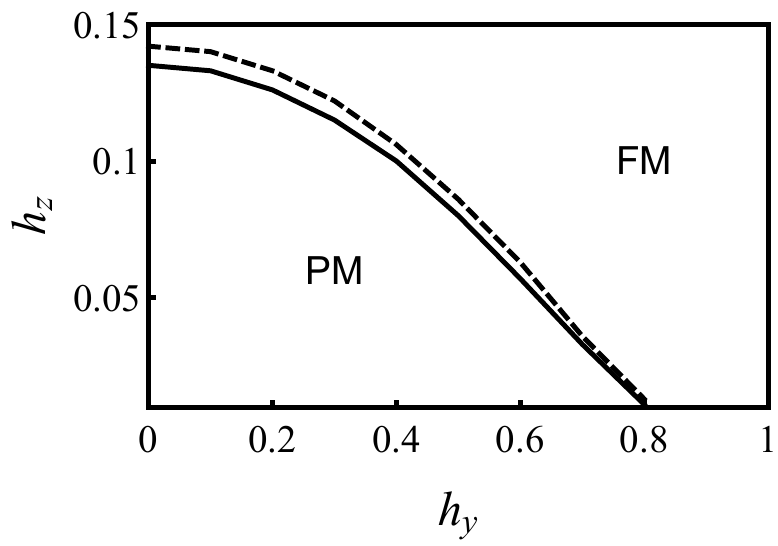}
	\caption{The phase diagram of Hamiltonian~(\ref{eq:hamiltonian}) obtained from exact diagonalization of a chain of $L=10$ with periodic boundary condition. The solid line denotes the second-order Yang-Lee critical line, and the dashed line is the first-order entanglement transition. We choose $J=0.4$, $J_2=0.1$ and $\Gamma = 1$.}
	\label{fig:phase_diagram}
\end{figure}

When $h_y=0$, Hamiltonian~(\ref{eq:hamiltonian}) belongs to the same universality class as the 2D classical Ising model, and the Yang-Lee singularity is realized at $h_z \neq 0$ in the paramagnetic phase with $\Gamma > J$~\cite{von:1991critical}. Since the ground state of Hamiltonian~(\ref{eq:hamiltonian}) is non-degenerate in the paramagnetic phase, its energy remains real upon increasing $h_z$ until the gap to the first excited state closes at the critical point, as shown in Fig.~\ref{fig:spectrum_transition}. When the gap reopens past the critical point, the doubly-degenerate ground states split in pairs along the  imaginary axis and acquire a magnetic order. When $h_y \neq 0$, it turns out that Hamiltonian~(\ref{eq:hamiltonian}) can be brought to the same form as when $h_y=0$ via a similarity transformation~\cite{deguchi2009exactly}
\bea \label{eq:similar}
	H' =  - \sum_{i} (J \sigma^z_i \sigma^z_{i+1}+J_2 \sigma^z_i \sigma^z_{i+2} + \widetilde{\Gamma} \sigma^x_i + i h_z \sigma_i^z ),
\eea
where $H$ and $H^\prime$ are connected by an operator $\rho$, i.e., $H^\prime=\rho H \rho^{-1}$, and $\widetilde{\Gamma}=\sqrt{\Gamma^2-h_y^2}$, provided that $|h_y|<|\Gamma|$. 
Since Hamiltonian~(\ref{eq:similar}) has a $\mathcal{P}\mathcal{T}$ symmetry $\mathcal{P}\mathcal{T} = \prod_{i=1}^L \sigma_i^x \mathcal{K}$, Hamiltonian~(\ref{eq:hamiltonian}) also has a generalized $\mathcal{P}\mathcal{T}$ symmetry: $\mathcal{P}\mathcal{T} = \rho \prod_{i=1}^L \sigma_i^x \mathcal{K}$, where $\mathcal K$ denotes complex conjugation.
Thus, a nonzero $h_y$ simply attenuates the effective strength of the transverse field $\Gamma$, and
the Yang-Lee singularity persists for a range of nonzero $h_y$. The phase diagram of Hamiltonian~(\ref{eq:hamiltonian}) is shown in Fig.~\ref{fig:phase_diagram}. Qualitatively, this phase diagram can be obtained using a mean-field theory of Hamiltonian~(\ref{eq:hamiltonian})~\cite{SM}.
Since there is only one relevant direction for the Yang-Lee critical point, we shall hereafter fix $h_y=0$ and vary $h_z$.

{\it Entanglement transition}.-- We are interested in the entanglement properties of the long-time steady state evolved under Hamiltonian~(\ref{eq:hamiltonian})
\begin{equation}
    |\psi(t) \rangle = \frac{e^{-i H t} |\psi_0 \rangle}{||e^{-i H t} |\psi_0 \rangle||},
\end{equation}
for $t\gg 1$, where $|\psi_0\rangle$ is an unentangled initial state. In the long-time limit, $|\psi(t)\rangle$ is dominated by the eigenstate of $H$ whose imaginary part of the eigenenergy ${\rm Im}(E)$ is the largest. It is thus possible to infer the entanglement property of the long-time steady state from a {\it single} eigenstate with the largest ${\rm Im}(E)$. In Fig.~\ref{fig:transition_imag0}, we plot ${\rm Im}(E)$ for the ground state and the eigenstate with the largest ${\rm Im}(E)$, respectively, as $h_z$ increases. Remarkably, we find a level crossing in ${\rm Im}(E)$ shortly after the Yang-Lee edge singularity, when the ground state takes over to be the one with the largest ${\rm Im}(E)$. Due to $\mathcal{PT}$ symmetry, two eigenvalues must coalesce before wandering off the real axis in pairs. One thus expects that, prior to this level crossing, eigenstates that are most likely to develop a large ${\rm Im}(E)$ and hence control the steady state are those located near the middle of the spectrum, where level spacings are the smallest and scale as $2^{-L}$. Since these eigenstates are inherited from the excited states of the chaotic Hermitian Hamiltonian $H_1$, we expect them to continue exhibiting a volume-law entanglement entropy upon turning on $H_2$, as long as the non-Hermitian part is not too large. The volume-law entanglement scaling of such eigenstates in the presence of $H_2$ is numerically demonstrated in Figs.~\ref{fig:transition_ee}~\&~\ref{fig:circuit}(b) (see below). On the other hand, the ground state is close to a product state with low entanglement obeying an area-law scaling with the subsystem size. Therefore, such a level crossing gives rise to a first-order entanglement transition in the long-time steady state, across which the entanglement jumps discontinuously from a volume-law to an area-law scaling. In Fig.~\ref{fig:transition_order0}, we show that the half-chain entanglement entropy of the maximal ${\rm Im}(E)$ eigenstate indeed exhibits a discontinuous jump at the level crossing. The scaling of the entanglement entropy before and after the jump with subsystem sizes shown in Fig.~\ref{fig:transition_ee} also confirms the volume-to-area-law nature of the transition~\cite{nonthermal}.

\begin{figure}[t]
    \centering
\subfigure[]{\label{fig:transition_ee}
    \includegraphics[width=0.215\textwidth]{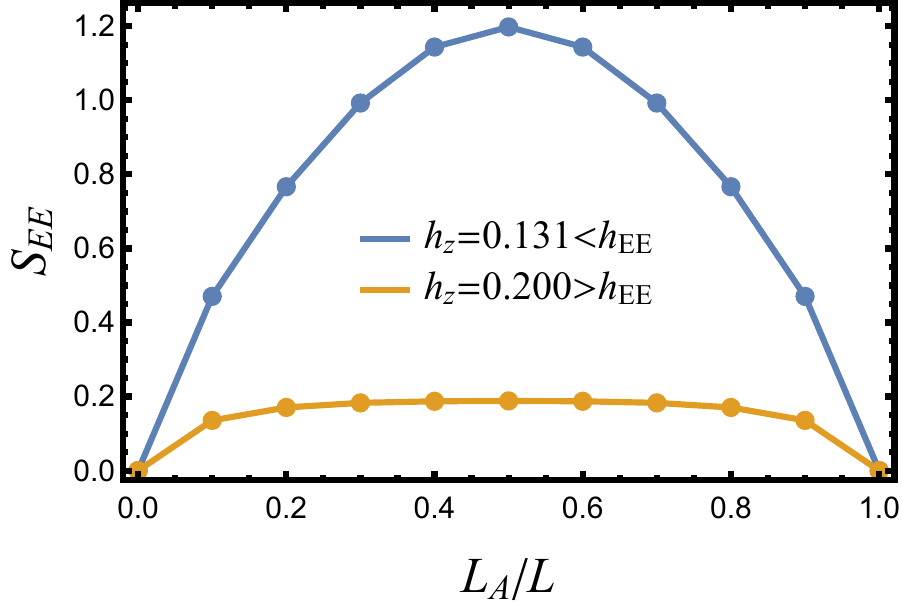}} \quad
\subfigure[]{ \label{fig:e-h_scaling}
	\includegraphics[width=0.22\textwidth]{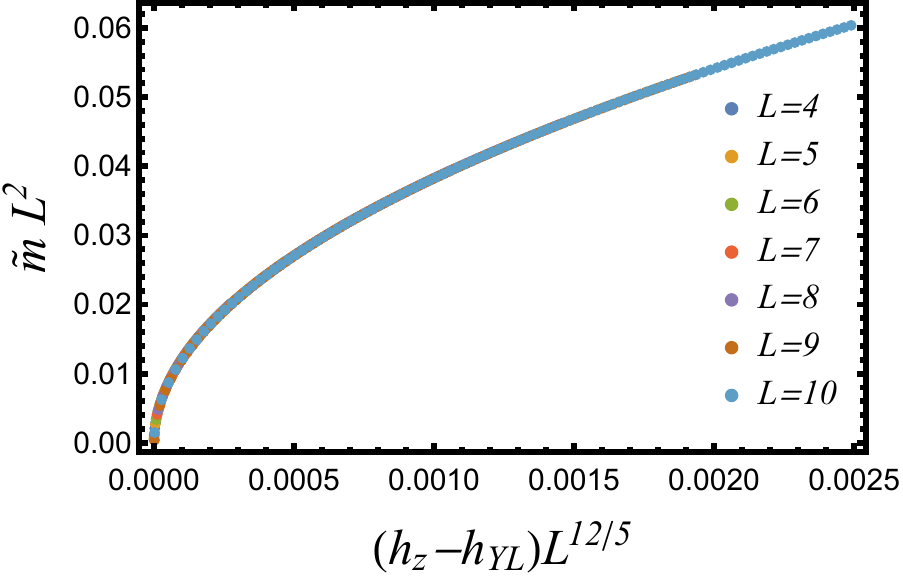}}
	\caption{(a) Scaling of the entanglement entropy as a function of subsystem sizes before and after the level crossing.
	(b) Data collapse of $\widetilde{m}$ as a function of $(h_z-h_{YL})$ for different system sizes. The choices of parameters are the same as in Fig.~\ref{fig:2}.}
\end{figure}

This first-order entanglement transition, although seemingly coincidental, is in fact {\it guaranteed} by the Yang-Lee singularity. First of all, the ground state energy can acquire a non-zero imaginary part solely due to the existence of a critical point where the ground state becomes degenerate, as required by $\mathcal{PT}$ symmetry. Secondly, the development of ferromagnetic ordering in the ground state past the Yang-Lee singularity guarantees that the ground state will eventually have the largest ${\rm Im}(E)$. An observation of Hamiltonian~(\ref{eq:hamiltonian}) yields ${\rm Im}(E) \propto \widetilde{m} L$ for an eigenstate, where $\widetilde{m} \equiv \frac{1}{L}\sum_i \langle \sigma_i^z\rangle$ is the average magnetization of this eigenstate~\cite{footnote}. One thus expects that the ferromagnetically ordered ground state has the largest magnetization and hence its ${\rm Im}(E)$ must dominate over states near the middle of the spectrum in the thermodynamic limit. Therefore, although the long-time steady state by itself is blind to the critical point, the very existence of which in fact triggers a subsequent level crossing, when the steady state switches character from a highly entangled state in the middle of the spectrum to an ordered ground state with low entanglement.

To further show that the first-order entanglement transition happens at a finite distance past the critical point in the thermodynamic limit, we employ a finite-size scaling analysis of the onset of ${\rm Im}(E)$ in the vicinity of the critical point. Since the Yang-Lee critical point has a dynamical critical exponent $z=1$, it is natural to expect that the (imaginary) energy density $\widetilde{m} \propto {\rm Im}(E)/L$ should satisfy the following scaling form
\bea
    \widetilde{m} = L^{-2}  f_{\widetilde{m}} \left( (h_z-h_{YL}) L^{\frac{d+2-\eta}2} \right), \quad h_z > h_{YL},
\eea
where $f_{\widetilde{m}}$ is a universal scaling function with $f_{\widetilde{m}}(0)=0$, and $\eta$ is the anomalous dimension. In order to have a sensible scaling form in the thermodynamic limit, the scaling function must satisfy $f_{\widetilde{m}}(x)\sim x^{\frac{4}{d+2-\eta}}$ as $x \rightarrow \infty$, yielding $ {\widetilde{m}} \sim  (h_z-h_{YL})^{\frac{4}{d+2-\eta}}$ in the thermodynamic limit. For the (1+1)-D Yang-Lee singularity that we focus on here, $d=2$, and the corresponding non-unitary CFT data give $\eta = -4/5$~\cite{Cardy:1985conformal, SM}. The scaling form thus becomes ${\widetilde{m}} L^2 = f_{{\widetilde{m}}}\left( (h_z-h_{YL}) L^{12/5} \right)$. This relation is demonstrated perfectly in Fig.~\ref{fig:e-h_scaling}. 
We thus obtain the following universal scaling form of $\widetilde{m}$ of the ground state near the Yang-Lee singularity in the thermodynamic limit (with $\lambda>0$):
\bea \label{eq:scale2}
	{\widetilde{m}} \approx \lambda (h_z-h_{YL})^{5/6}, \quad  h_z>h_{YL}.
\eea
Eq.~(\ref{eq:scale2}) implies that ${\widetilde{m}}$ continuously increases from zero in the thermodynamic limit, and hence the first-order entanglement transition must happen at a finite distance past the critical point when ${\widetilde{m}}$ of the ground state supersedes that of the previously dominating eigenstate.
If one instead starts from the ferromagnetic phase of Hamiltonian~(\ref{eq:hamiltonian}), this entanglement transition is absent. Since the ground state is twofold degenerate to begin with, an infinitesimal $h_z$ will immediately drive the steady state to an area-law phase.

We remark that the mechanism underlying the entanglement transition as being triggered by a critical point is different from that in recently studied non-Hermitian systems where a critical point is absent~\cite{gopalakrishnan2020entanglement}. In the Supplemental Materials, we show that the ground state of the model studied therein remains non-degenerate with real energy at all times, due to the absence of a quantum phase transition~\cite{SM}. Therefore, a level crossing is not guaranteed, and the entanglement entropy of the steady state evolves continuously once the spectrum becomes complex.

\begin{figure}
    \centering
\subfigure[]{\label{fig:circuit_transition}
    \includegraphics[width=0.23\textwidth]{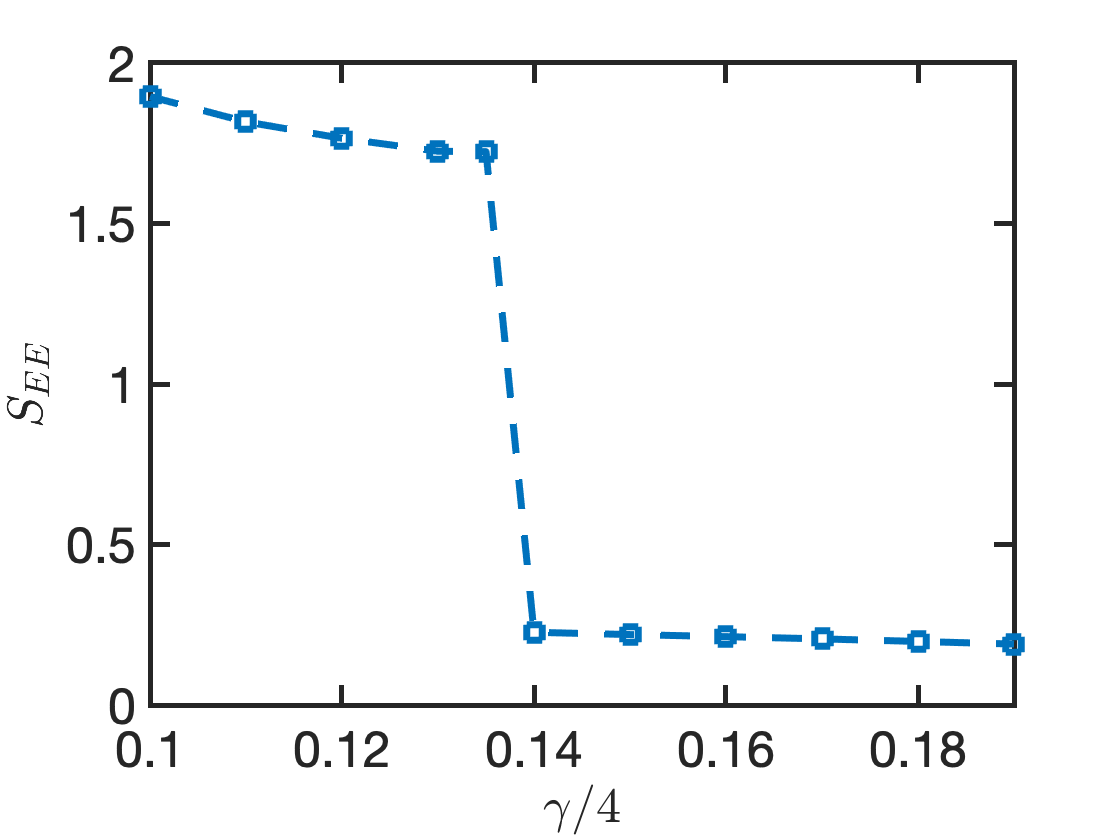}}
\subfigure[]{\label{fig:circuit_ee}
    \includegraphics[width=0.23\textwidth]{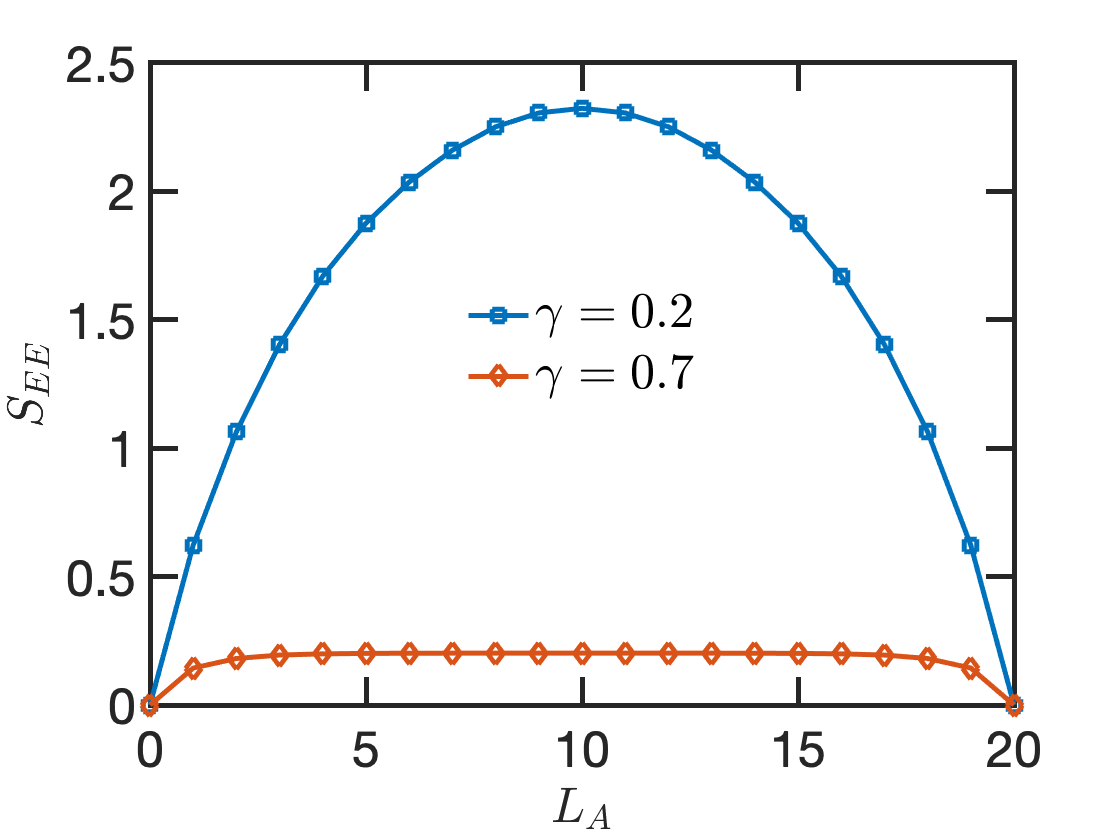}} 
    \caption{
    (a) Entanglement entropy of the steady state under the hybrid circuit evolution~(\ref{eq:circuit}) as a function of $\gamma$ for $L=16$.
    (b) Scaling of the steady state entanglement entropy at measurement rates $\gamma < \gamma_{EE}$ and $\gamma > \gamma_{EE}$, respectively, for $L=20$. We choose $J=0.4$, $J_2=0.1$, $\Gamma=1$, $\tau = 0.1$, and periodic boundary condition.}
    \label{fig:circuit}
\end{figure}

{\it Forced-measurement induced entanglement transition}.-- It is useful to connect the time evolution under the non-Hermitian Hamiltonian~(\ref{eq:hamiltonian}) to a system undergoing repeated weak measurements and post-selections. Consider the following circuit. In each time step of duration $\tau$, the circuit consists of a unitary time evolution $U = e^{-\ii H_0 \tau }$ with $H_0 = - \sum_{i=1}^L (J \sigma^z_i \sigma^z_{i+1} + J_2 \sigma^z_i \sigma^z_{i+2} + \Gamma \sigma^x_i)$, followed by weak measurements corresponding to the following set of Kraus operators 
\bea
    M_0^{(i)} = 1 - \left(1-\sqrt{1-\gamma \tau} \right) \Pi_i \quad    M_z^{(i)} = \sqrt{\gamma \tau} \Pi_i,
\eea
where $\Pi_i = \frac12 (\sigma^z_i + 1)$ is a projector to the spin-up state at site $i$. If the post-selection is conditioned on $M_0^{(i)} $, the time evolution with $M_0 = \otimes_{i=1}^L M_0^{(i)}$
\begin{equation}
    \big|\psi(N\tau) \big\rangle = \frac{(M_0 U)^N |\psi_0\rangle}{||(M_0 U)^N |\psi_0\rangle||}
    \label{eq:circuit}
\end{equation}
in the limit $\gamma \tau \ll 1$ is then precisely generated by the non-Hermitian Hamiltonian~(\ref{eq:hamiltonian}) with $ h_z = \gamma/4 $ and $ h_y = 0$. In general, one can rotate the spin polarization direction in the $y$-$z$ plane along which measurements are performed so as to realize Hamiltonian~(\ref{eq:hamiltonian}) with arbitrary $h_y$ and $h_z$. The entanglement transition in this context is also known as a forced-measurement-induced phase transition~\cite{nahum2020measurement}. Namely, when the measurement rate is finite but smaller than a threshold $\gamma < \gamma_{EE}$, the hybrid circuit is able to evolve an unentangled inital state to a final state with volume-law entanglement; whereas when the measurement rate is large $\gamma > \gamma_{EE}$, the time-evolved state remains area-law entangled. We simulate the time evolution~(\ref{eq:circuit}) for system size up to $L=20$ using Krylov space time evolution method~\cite{luitz2017ergodic}, and compute the entanglement entropy for different measurement rates, as shown in Fig.~\ref{fig:circuit}.
We indeed find a similar entanglement transition in the steady state from a volume-law to an area-law scaling, as the measurement rate increases~\cite{SM}. Such a forced-measurement-induced entanglement transition can now be elegantly accounted for by a Yang-Lee edge singularity triggered level crossing in the eigenspectrum of the corresponding non-Hermitian Hamiltonian~\cite{SM}. Since this transition is first order, it is distinct from the continuous transitions driven by weak measurements where there is intrinsic randomness in the outcomes~\cite{PhysRevB.100.134306}.

{\it Spin-1 Blume-Capel model}.-- To demonstrate that this mechanism for entanglement transition extends beyond the simple Hamiltonian~(\ref{eq:hamiltonian}),
we now show another quantum spin chain realization of the Yang-Lee edge singularity, where the critical point also triggers a subsequent entanglement transition. Consider the quantum spin-1 Blume-Capel model described by the Hamiltonian
\begin{equation}
    H = \sum_i \left[ \alpha (S_i^z)^2 + \beta S_i^x - S_i^z S_{i+1}^z - ih S_i^z \right],
\label{eq:blume}
\end{equation}
where $S_i^{x,y,z}$ are $3\times 3$ spin-1 matrices. This model has a rich phase diagram, as shown in Ref.~\cite{gehlen1994non}. In the absence of an imaginary field $h=0$, Hamiltonian~(\ref{eq:blume}) has an ordered phase with broken $\mathbb{Z}_2$ symmetry and a disordered phase separated by a single critical curve starting at $\alpha=1$ and $\beta=0$, and moving towards smaller values of $\alpha$ upon increasing $\beta$. In the disordered phase, further turning on the imaginary field $h$ drives a continuous transition with $c=-22/5$ corresponding to the Yang-Lee universality class. 

\begin{figure}[t]
    \centering
    \includegraphics[width=.45\textwidth]{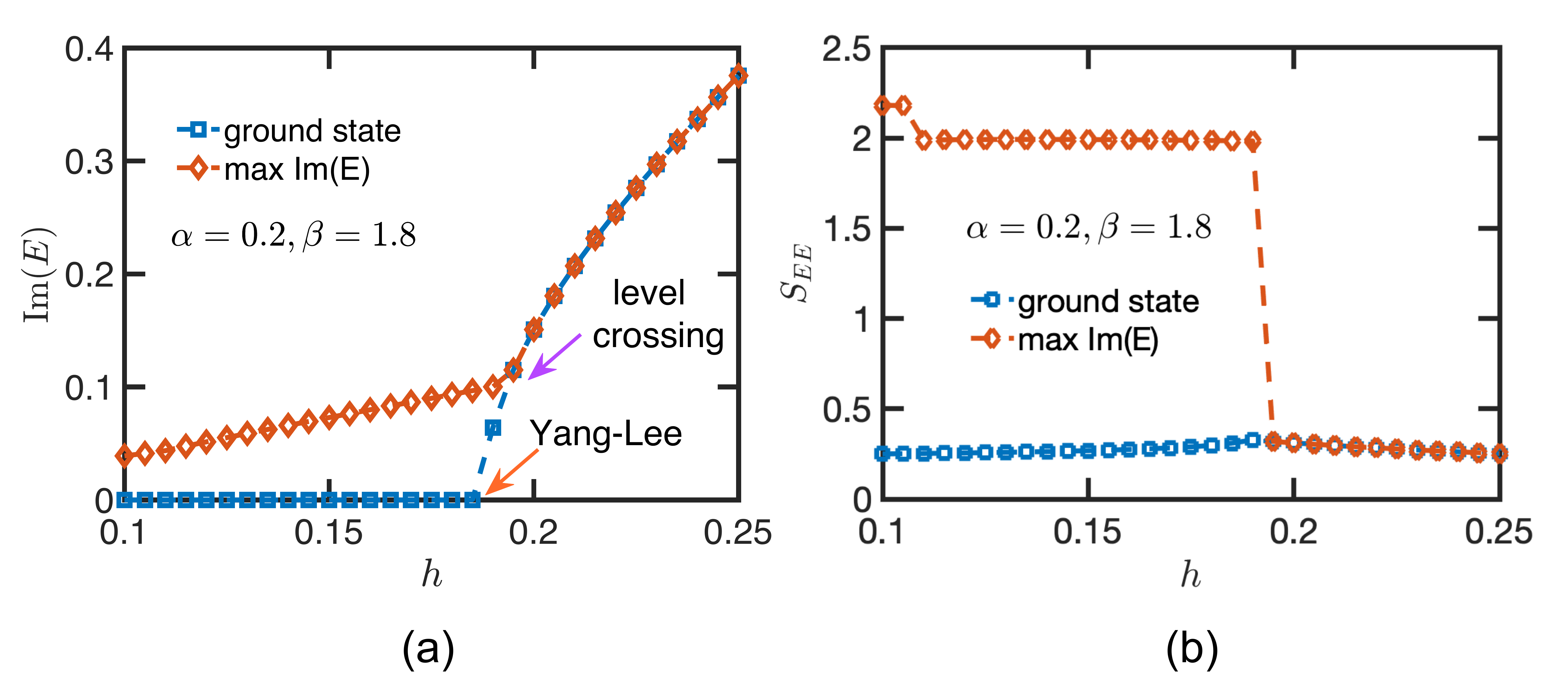}
    \caption{(a) Imaginary part of the eigenvalues for the ground state and eigenstate with the largest ${\rm Im}(E)$ of Hamiltonian~(\ref{eq:blume}). (b) Entanglement entropy of the same sets of states as in (a). We choose $\alpha=0.2, \beta=1.8$ and system size $L=6$ with periodic boundary condition.}
    \label{fig:blume}
\end{figure}

In Fig.~\ref{fig:blume}(a), we plot the imaginary part of the eigenvalues for the ground state and eigenstate with the largest ${\rm Im}(E)$ of Hamiltonian~(\ref{eq:blume}) as a function of $h$. Similar to Hamiltonian~(\ref{eq:hamiltonian}), here we also find a level crossing slightly past the Yang-Lee singularity, after which the ground state becomes the one with the largest ${\rm Im}(E)$. We thus expect that the steady state entanglement entropy will also exhibit a discontinous jump from a volume-law scaling to an area-law scaling, as is confirmed numerically in Fig.~\ref{fig:blume}(b). Since the ground state develops a large magnetization $\langle S^z \rangle$ after transitioning to the ordered phase, we argue that the imaginary part of the ground state energy, which is proportional to $\langle S^z \rangle L$, must dominate over all other eigenstates. Therefore, this level crossing, and hence the first-order entanglement transition, persists in the thermodynamic limit.

So far, we have been focusing on non-Hermitian Hamiltonians with $\mathcal{PT}$ symmetry as a cleanest exemplification of our general idea.
However, we remark that $\mathcal{PT}$ symmetry is {\it not} a necessary condition for the entanglement transition to happen, as long as the system harbors a critical point. In the Supplemental Materials, we give an example of the quantum three-state Potts model with an imaginary field, which does not have $\mathcal{PT}$ symmetry but nevertheless possesses the Yang-Lee edge singularity~\cite{SM}. We find that the same story holds in this case as well.

{\it Concluding remarks}.-- In this work, we demonstrate a new mechanism where a class of non-Hermitian Hamiltonians realizing the Yang-Lee edge singularity further triggers a first-order entanglement transition in the long-time steady state. This entanglement phase transition can also be understood in terms of the purification dynamics~\cite{PhysRevLett.125.070606, PhysRevX.10.041020, SM}. 
 Notice that the purification rate is determined by the gap between the largest and second largest $\mbox{Im}(E)$. In the area-law phase, the entropy of an initially mixed density matrix decays at a finite constant rate and the purification time scales as $\log L$. In contrast, in the volume-law phase, the purification time is much longer, presumably exponential in the system size.
The  mechanism  for entanglement  transitions uncovered in this work also provides new insight on non-unitary dynamics from the perspective of quantum trajectories. If post-selections are removed or some randomness is introduced in our model, we expect that this first-order transition will be rounded to a continuous one~\cite{aizenman1989rounding, cardy1997critical}. We leave a detailed study on this for future work.

{\it Acknowledgments}.-- We thank Michael Gullans for useful comments on the manuscript, and Sarang Gopalakrishnan and Thomas Iadecola for helpful discussions.
S.-K. J. is supported by the Simons Foundation via the It From Qubit Collaboration. Z.-C. Y. acknowledges funding by the DoE ASCR Accelerated Research in Quantum Computing program (award No. DE-SC0020312), U.S. Department of Energy Award No. DE-SC0019449, DoE ASCR Quantum Testbed Pathfinder program (award No. DE-SC0019040), NSF PFCQC program, AFOSR, ARO MURI, AFOSR MURI, and NSF PFC at JQI.  Z.-C. Y. is also supported by MURI ONR N00014-20-1-2325, MURI AFOSR, FA9550-19-1-0399, and Simons Foundation. Part of the numerical calculations were performed on the Boston University Shared Computing Cluster, which is administered by Boston University Research Computing Services.

\bibliography{reference}

\begin{thebibliography}{47}
\bibitem [{foo({\natexlab{a}})}]{footnote2}%
  \BibitemOpen
  \href@noop {} {} ({\natexlab{a}}),\ \bibinfo {note} {notice that the states
  here are not really steady, since the wavefunction is not normalized due to
  the non-unitary evolution. Nonetheless, here we use the name to refer to
  eigenstate with the largest imaginary eigenvalue, so that it dominates the
  wavefunction in the long-time limit.}\BibitemShut {Stop}%
\bibitem [{non()}]{nonthermal}%
  \BibitemOpen
  \href@noop {} {}\bibinfo {note} {Due to a finite non-Hermitian part, the
  volume law coefficient is much less than the thermal value at
  $h_z=0$.}\BibitemShut {Stop}%
\bibitem [{foo({\natexlab{b}})}]{footnote}%
  \BibitemOpen
  \href@noop {} {} ({\natexlab{b}}),\ \bibinfo {note} {notice that the
  magnetization referred to here as being a $\sigma^z$ sandwiched between two
  right eigenvectors is {\it not} the order parameter associated with the
  corresponding field theory, which is a $\sigma^z$ sandwiched between a left
  and a right eigenvector and satisfies the scaling relation $m \sim
  (h-h_c)^{1/\delta}$. We draw a distinction by adding a tilde to the quantity
  we defined here: $\widetilde{m}$.}\BibitemShut {Stop}%
\end{thebibliography}

\newpage
\onecolumngrid
\appendix

\subsection*{Supplemental Material for "Yang-Lee edge singularity triggered entanglement transition"}

\section{Review of the Yang-Lee edge singularity}

In this section we briefly review the effective $i \phi^3$ field theory describing the Yang-Lee edge singularity~\cite{Fisher:1978yang}. The coarse-grained Hamiltonian of an Ising model in the presence of an imaginary magnetic field is given by the well-known $\phi^4$ theory
\bea
	H = \frac{r}2 m^2 + \frac{\lambda}{4!} m^4 - i h m, \quad h \in \mathbb{R}, 
\eea
where $m$ is the order parameter. In the paramagnetic phase $r>0$, and we can shift the order parameter $m$ by a constant:  $m =  i\sqrt{2r/\lambda } + \phi$, such that the quadratic term vanishes. The Hamiltonian now takes the form
\bea
	 H=-i (h-h_0) \phi + i w \phi^3,
\eea
where $h_0 = \frac{2\sqrt2 r^{3/2}}{3 \sqrt \lambda}$, $w = \frac{\sqrt{r \lambda}}{3\sqrt 2},$ and higher-order terms that are irrelevant at low energies are neglected. The mean-field solution is
\bea
	h-h_0 = 3 w \phi^2, \quad \phi \sim (h-h_0)^{1/2},
\eea
from which one can read off the mean-field exponent $\delta = 2$. 

An interesting property of this theory is that there is only one independent exponent. Namely, the correlation length exponent $\nu$ is related to the anomalous dimension via
\bea
	\nu = \frac2{d+2 - \eta}.
\eea
This is because the tuning parameter for the transition is the external field $h$, and hence the exponent $\nu$ is given by the inverse of the renormalization group (RG) eigenvalue of the scaling variable $h$.
An intuitive understanding of this phenomena is that in order to have an effective theory for a physical fluctuation field, one can implicitly tune the imaginary part of this field, which relates the field tuning to the mass tuning. Other critical exponents are also listed below for completeness
\bea
	\alpha = 2 - \frac{2d}{d+2-\eta}, \quad \beta = \frac{d-2 + \eta}{d+2 + \eta}, \quad \gamma = \frac{4 - 2\eta}{d +2 - \eta}, \quad \delta = \frac{d+2-\eta}{d-2 + \eta}.
\eea

Going beyond mean-field, one can calculate the one-loop correction. At the critical point $h=h_0$, the effective action becomes
\bea \label{eq:effective}
	S = \frac12 \int d^d {\bf x}( \nabla \phi)^2 + i w \int d^d {\bf x} \phi^3,
\eea
where $\nabla \equiv \partial_{\bf x}$. It is straightforward to get the RG equations
\bea
	\eta &=& 18 \left(  \frac{4}d - 1 \right) \frac{S_d}{(2\pi)^d} w^2, \\
	\frac{d w}{d l} &=& \frac{6-d}2 w - \left( 27 \big( \frac4d-1 \big)+36 \right) \frac{S_d}{(2\pi)^d} w^3,
\eea
where $S_d$ is the surface area of the unit ($d-1$)-sphere. In principle the RG equation is valid for $\epsilon = 6- d \ll 1$. The stable fixed point is 
\bea
	(w^{*})^2 = \frac{\epsilon}{54} \frac{(2\pi)^d}{S_d}, \quad \eta = - \frac{\epsilon}9.
\eea
For a scalar field theory, the unitarity bound is $\Delta \ge \frac{d-2}2$~\cite{Poland:2019conformal}. The negative anomalous dimension $\eta = - \frac{\epsilon}9$ indicates that the theory is non-unitary. In particular, in (1+1) dimension it is known that the Yang-Lee edge singularity is described by a non-unitary CFT with central charge $c= -22/5$, corresponding to the minimal model series $\mathcal{M}_{2,5}$~\cite{Cardy:1985conformal}. The anomalous dimension in this case is $\eta = -4/5$.

The Yang-Lee edge singularity exists also in (0+1) dimension, where it is nothing but a finite matrix theory. One can illustrate this using the following $2\times 2$ matrix
\bea \label{eq:matrix}
	H = h_{YL} \sigma^x + i h \sigma^z,
\eea
with ground state energy
\begin{equation}
 E_0 = - \sqrt{h_{YL}^2-h^2},
\end{equation}
and (unnormalized) right and left eigenvectors
\begin{eqnarray}
	 |\psi_{0,r} \rangle &=& \left(i h/h_{YL} - \sqrt{1-(h/h_{YL})^2} , \ 1\right)^T, \\
	 |\psi_{0,l} \rangle &=& \left(-i h/h_{YL} - \sqrt{1-(h/h_{YL})^2} ,\ 1\right)^T,
\end{eqnarray}
such that 
\begin{equation}
H |\psi_{0,r} \rangle = E_0 |\psi_{0,r} \rangle, \quad H^\dag |\psi_{0,l} \rangle = E_0 |\psi_{0,l} \rangle. 
\end{equation}
The ground state energy $E_0$ implies that $\mathcal{PT}$ symmetry is broken if $h^2> h_{YL}^2$. One can also obtain the ground state magnetization
\bea \label{eq:matrix_magnetization}
	m = \frac{\langle \psi_{0,l} | \sigma^z | \psi_{0,r} \rangle}{\langle \psi_{0,l} | \psi_{0,r} \rangle } = - i \frac{h}{\sqrt{h_{YL}^2 - h^2}}, \quad
	m \sim (h-h_{YL})^{-1/2},
\eea
where the second equation leads to the critical exponent $\delta = -2$ and $\eta = -1$ for (0+1) dimensional Yang-Lee edge singularity. Notice that the magnetization as it appears in the field theory is a $\sigma^z$ sandwiched between a left and a right eigenvector, and is thus distinct from the quantity $\widetilde{m}$ discussed in the main text.

The lattice Hamiltonian we study in this work takes the following form:
\begin{equation}
H = -\sum_i (J \sigma_i^z \sigma_{i+1}^z + J_2 \sigma_i^z \sigma_{i+2}^z + \widetilde{\Gamma} \sigma_i^x + i h_z \sigma_i^z).
\label{eq:model}
\end{equation}
The partition function of the quantum Hamiltonian~(\ref{eq:model}) is given by $\mathcal{Z} = {\rm Tr} \ {\rm exp}(-\beta H)$. Following the standard procedure of quantum-classical mapping~\cite{sachdev2011quantum}, we split the imaginary time direction into $M$ slices with lattice spacing $a$: $\beta \equiv M a$. In the scaling limit $a\rightarrow 0$, $M \rightarrow \infty$, $\beta \rightarrow \infty$, the partition function can be written as:
\begin{equation}
\mathcal{Z} = \sum_{\{m_{i,j}\}} {\rm exp} \left[ \sum_{i,j} \left(J a \ m_{i,j} m_{i+1,j} + J_2 a \ m_{i,j} m_{i+2,j} + J_{\perp} \ m_{i,j} m_{i,j+1} + i h_z a \  m_{i,j} \right) \right],
\end{equation}
where the coupling along the imaginary time direction is given by ${\rm exp}(-2J_{\perp}) ={\rm tanh}(\widetilde{\Gamma}a)$. Therefore, the partition function of the quantum Hamiltonian maps to that of a \textit{classical} Ising model in two dimensions, in the presence of an imaginary longitudinal field. We thus expect that the phase transition driven by $ih_z$ in Hamiltonian~(\ref{eq:model}) studied in the main text should belong to the same universality class as the Yang-Lee singularity of the classical Ising model.

\section{Mean-field theory of Hamiltonian~(2) }

We now present a Weiss mean-field theory of Hamiltonian~(2). Let us first decouple the interaction term as
\bea
	\sigma_i^z \sigma_j^z = (\sigma^z_i + \sigma^z_j ) m - m^2,
\eea
where $ m = \langle \sigma^z \rangle$, so that the Hamiltonian reduces to a single spin in an external magnetic field
\bea 
	H = \sum_i \left[ (-2 \widetilde{J} m - i h_z) \sigma_i^z -\Gamma \sigma_i^x - i h_y \sigma_i^y + J m^2 \right],
\eea
where $\widetilde J = J + J_2$. It is now straightforward to get the eigenvalues, and at zero temperature, we only need the lower eigenvalue
\bea
	E_0 = \widetilde J m^2 - \sqrt{ 4 {\widetilde J}^2 m^2 - 4 i J h_z m - h_z^2 + \widetilde{\Gamma}^2 }, 
\eea
where $\widetilde{\Gamma} \equiv \sqrt{\Gamma^2 - h_y^2} $. So we find that the main effect of $h_y$ is to reduce the effective transverse field strength $\widetilde{\Gamma}$ when $h_y < \Gamma$. As we point out in the main text, there exists a similarity transformation that brings Hamiltonian~(2) to the same form as when $h_y=0$~\cite{deguchi2009exactly}
\bea
	H' =  - \sum_{i} (J \sigma^z_i \sigma^z_{i+1} + J_2 \sigma^z_i \sigma^z_{i+2}  + \widetilde{\Gamma} \sigma^x_i + i h_z \sigma^z ).
\eea
The same feature is also captured by our mean-field treatment above.

The mean-field equation is then $\partial_m E_0 = 0$, or equivalently the self-consistency equation $m = \langle \sigma^z \rangle$. Assuming that the system is near the critical point, $m = i m_0 + \phi$, with $\phi \ll 1$ (note that $m_0 \in \mathbb{R}$ does not need to be small), the phase boundary can be obtained by expanding the ground state energy as a function of $\phi$, and requiring the linear and quadratic terms to vanish
\bea
	\frac{h_z}{2 \widetilde J} = \pm \left[ \left( \frac{ \widetilde{\Gamma}}{2 \widetilde J} \right)^{2/3} - 1\right]^{3/2}.
\eea
At this phase boundary, if we include spatial and temporal fluctuations, the effective theory is nothing but the $i \phi^3$ theory shown in Eq.~(\ref{eq:effective}).

\begin{figure}[t]
	\centering
\subfigure[]{\label{fig:phase_diagram3}
    \includegraphics[width=0.35\textwidth]{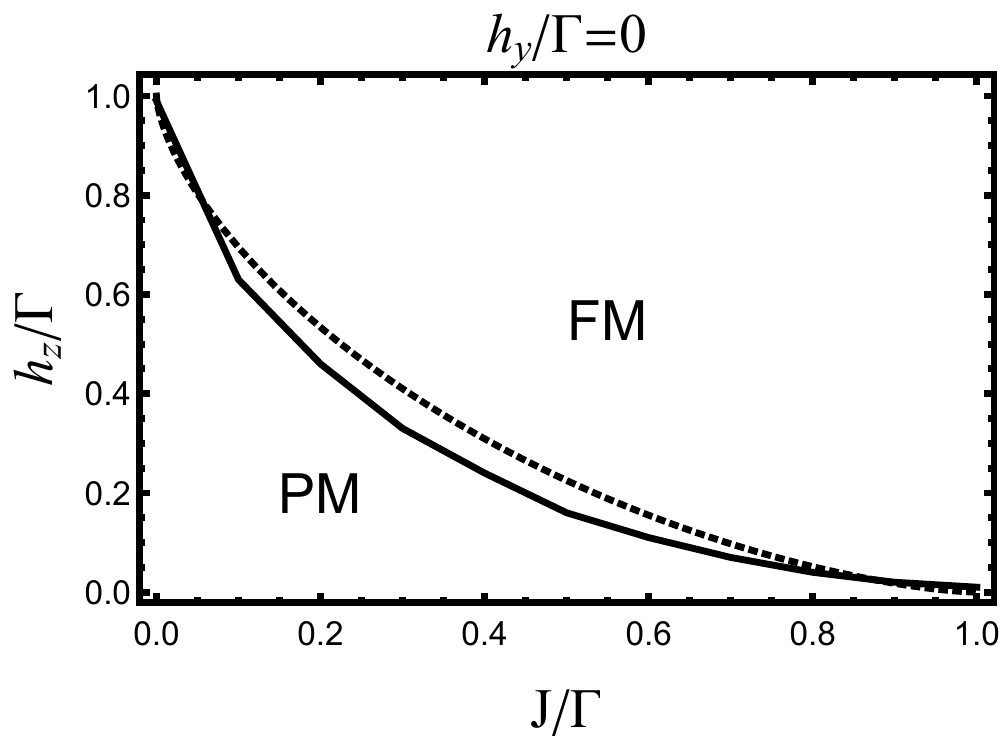}} \quad
\subfigure[]{\label{fig:phase_diagram2}
    \includegraphics[width=0.35\textwidth]{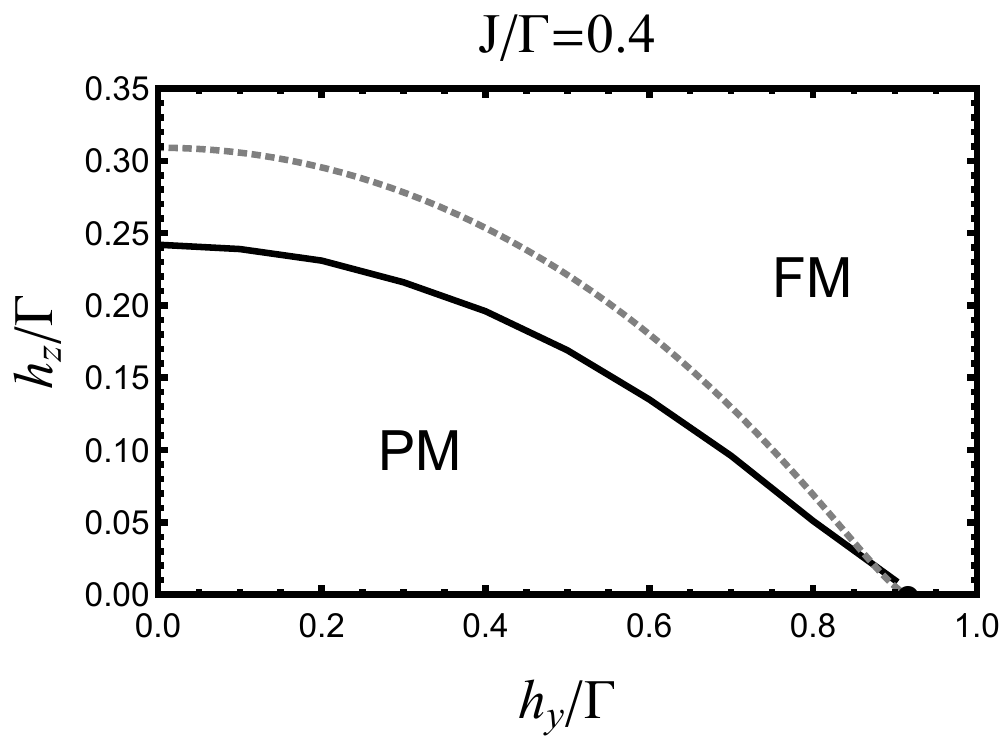} }
	\caption{The phase diagram of Hamiltonian~(2) in the main text for (a) $h_y/\Gamma = 0$ and (b) $J/\Gamma=0.4$. The solid lines are obtained from exact diagonalization of a chain of $L=8$ with periodic boundary condition, and the dotted lines are the mean-field phase boundary Eq.~(\ref{eq:phase_boundary}).}
\end{figure}

The Weiss mean-field theory usually overestimates the effect of interaction. For instance, it predicts that the conventional Ising transition at $J_2 = 0$ happens at $\Gamma = 2J$, whereas the true transition point is located at $\Gamma = J$. To simply account for this overestimate, we can regard the interaction in the Weiss mean-field theory as an effective interaction $ \widetilde{J}_{\rm eff} = \widetilde{J}/2$, and the "renormalized" phase boundary is given by
\bea \label{eq:phase_boundary}
	\frac{h_z}{\widetilde{J} } = \pm \left[ \left( \frac{ \widetilde{\Gamma}}{\widetilde{J}} \right)^{2/3} - 1\right]^{3/2}.
\eea
This agrees well with the exact phase boundary obtained using exact diagonalization, as shown in Fig.~\ref{fig:phase_diagram3} and Fig.~\ref{fig:phase_diagram2} for different choices of parameters.

\section{Scaling analysis of the Yang-Lee singularity in Hamiltonian~(2)}

\begin{figure}
	\centering
	\subfigure[]{ \label{fig:spectrum1}
		\includegraphics[width=0.35\textwidth]{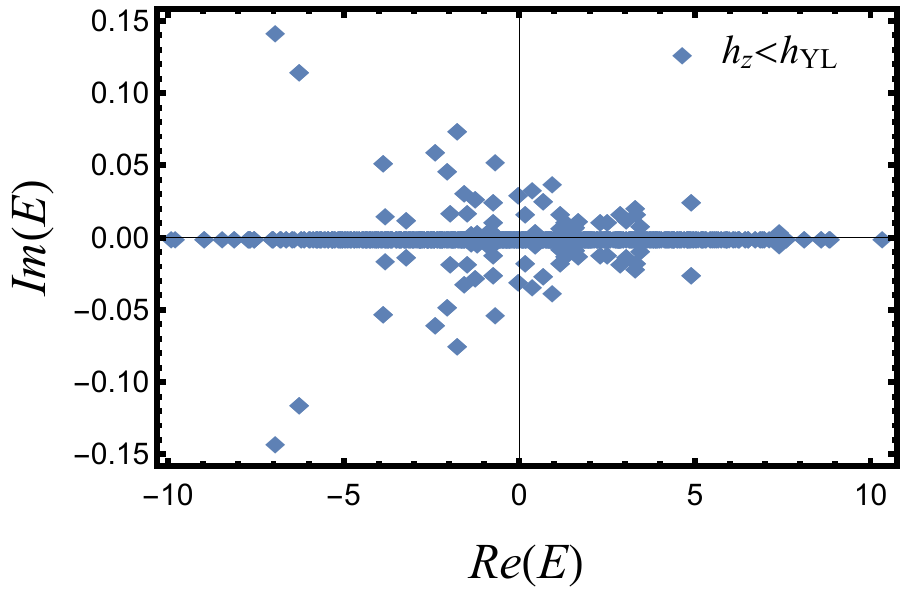}} \quad
	\subfigure[]{ \label{fig:spectrum2}
		\includegraphics[width=0.35\textwidth]{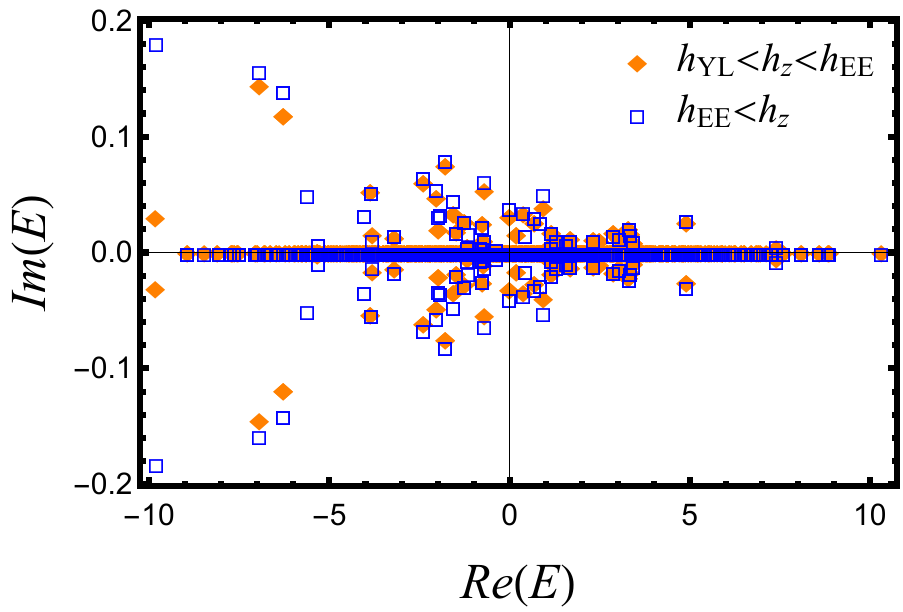}}
	\caption{The eigenenergies of Hamiltonian (2) before (a) and after (b) the Yang-Lee critical point. $h_{EE}$ denotes
		the entanglement transition point. We choose $J = 0.4$, $J_2 = 0.1$, $\Gamma= 1$, $h_y = 0$, and system size $L = 10$ with periodic
		boundary condition.}
\end{figure}

\begin{figure}[t]
	\centering
\subfigure[]{ \label{fig:m-h_single}
	\includegraphics[width=0.345\textwidth]{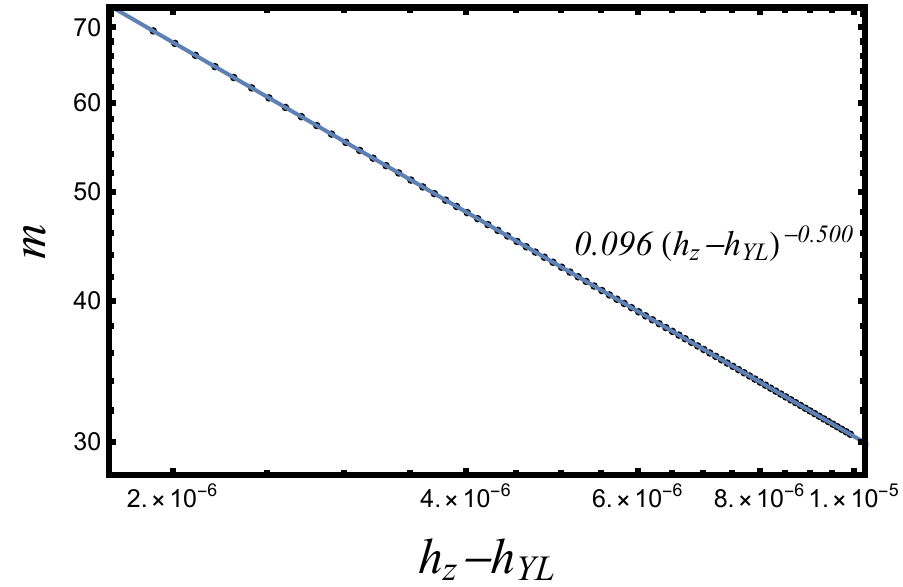}} \quad
\subfigure[]{ \label{fig:m-h_scaling}
	\includegraphics[width=0.35\textwidth]{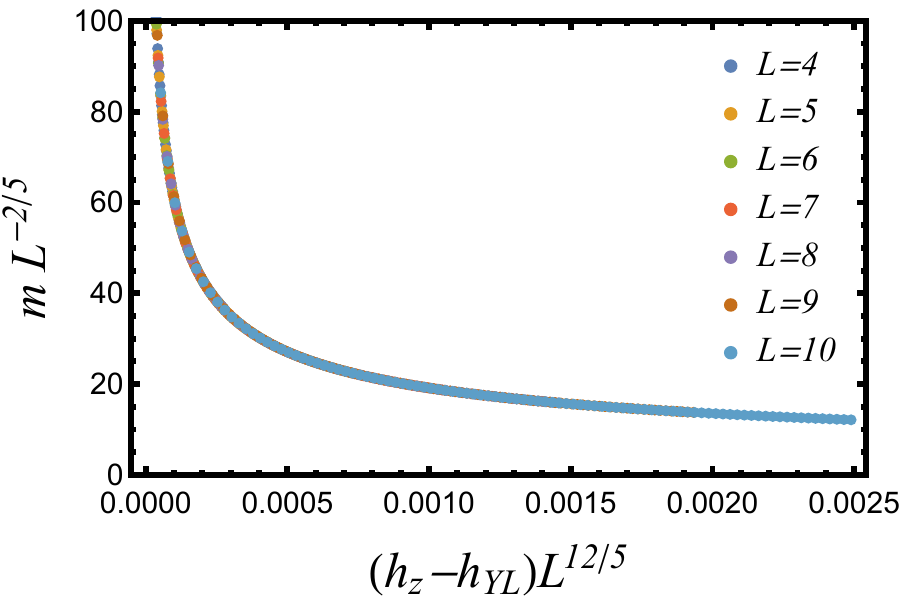}}
	\caption{(a) The log-log plot of the magnetization $m$ as a function of $(h_z-h_{YL})$ near the Yang-Lee edge singularity for a fixed system size $L=10$. (b) Data collapse of $m$ as a function of $(h_z-h_{YL})$ for various system sizes. The parameters are $J = 0.4, J_2 = 0.1, \Gamma = 1, h_y = 0$.}
\end{figure}

In this section, we first present exact diagonalization results of the eigenenergy spectrum of Hamiltonian~(2) before and after the Yang-Lee singularity $h_{YL}$ as well as the entanglement transition $h_{EE}$. This demonstrates the scenario illustrated in Fig.~1(a) of the main text.
We then provide additional finite-size scaling analysis of the Yang-Lee edge singularity realized in Hamiltonian~(2) in the main text. We focus on $h_y = 0$. For a fixed finite system of size $L$ and being sufficiently close to the critical point, the finite spatial extent of the system becomes irrelevant, and the transition is described by the (0+1) dimensional Yang-Lee edge singularity, where the magnetization scales as
\bea \label{eq:scaling1}
	m \sim (h_z-h_{YL})^{-1/2}, \quad h_z > h_{YL}
\eea
given by the simple matrix model~(\ref{eq:matrix_magnetization}). This scaling form is indeed confirmed numerically in Fig.~\ref{fig:m-h_single}.

The complete finite-size scaling function for the magnetization near the (1+1) dimensional Yang-Lee edge singularity takes the following form
\bea
	m = L^{- \frac{d-2+\eta}2} f_m\left( (h_z-h_{YL}) L^{\frac{d+2-\eta}2} \right), \quad h_z > h_{YL},
\eea
where $f_m$ is a universal scaling function with $f_m(0)=0$. In (1+1)D, $d=2, \eta = -4/5$, the scaling function is thus 
\begin{equation}
m L^{-2/5} = f_m\left( (h_z-h_{YL}) L^{12/5} \right).
\end{equation}
The above scaling form is confirmed numerically in Fig.~\ref{fig:m-h_scaling}, where the data obtained from different system sizes and parameters collapse to a single smooth function.

\section{Purification rate of Hamiltonian~(2)}

\begin{figure}[b]
	\centering
	\includegraphics[width=0.34\textwidth]{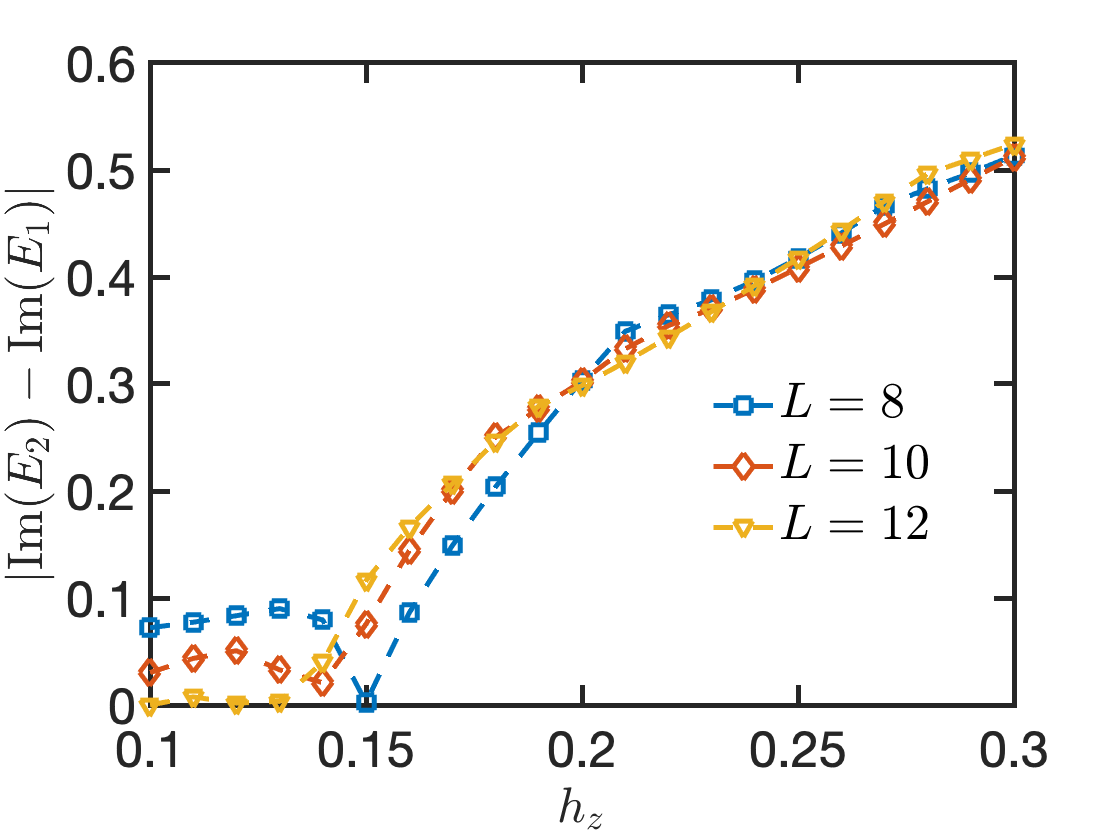}
	\caption{Purification rate, which is the gap between the largest and second largest ${\rm Im}(E)$, for Hamiltonian~(2) as a function of the imaginary longitudinal field. We fix $J=0.4$, $J_2=0.1$, $\Gamma=1$, and $h_y=0$.}
	\label{fig:gap}
\end{figure}

We now provide an interpretation of the entanglement transition in terms of the purification dynamics, using Hamiltonian~(2) as an example. In Fig.~\ref{fig:gap}, we plot the purification rate, i.e., the gap between the largest and second largest ${\rm Im}(E)$ as a function of the imaginary longitudinal field. We find that the purification rate in the volume-law phase is small, due to the small average level spacings near the middle of the spectrum, and decreases upon increasing system sizes. On the other hand, the purification rate starts increasing immediately past the entanglement transition point, and shows very little system-size dependence. This implies that an initially mixed state purifies in a much shorter (and perhaps system-size independent) timescale in the area-law phase than in the volume-law phase.

\section{Forced-measurement induced entanglement transition in the circuit model}

\begin{figure}[t]
	\centering
\subfigure[]{\includegraphics[width=0.34\textwidth]{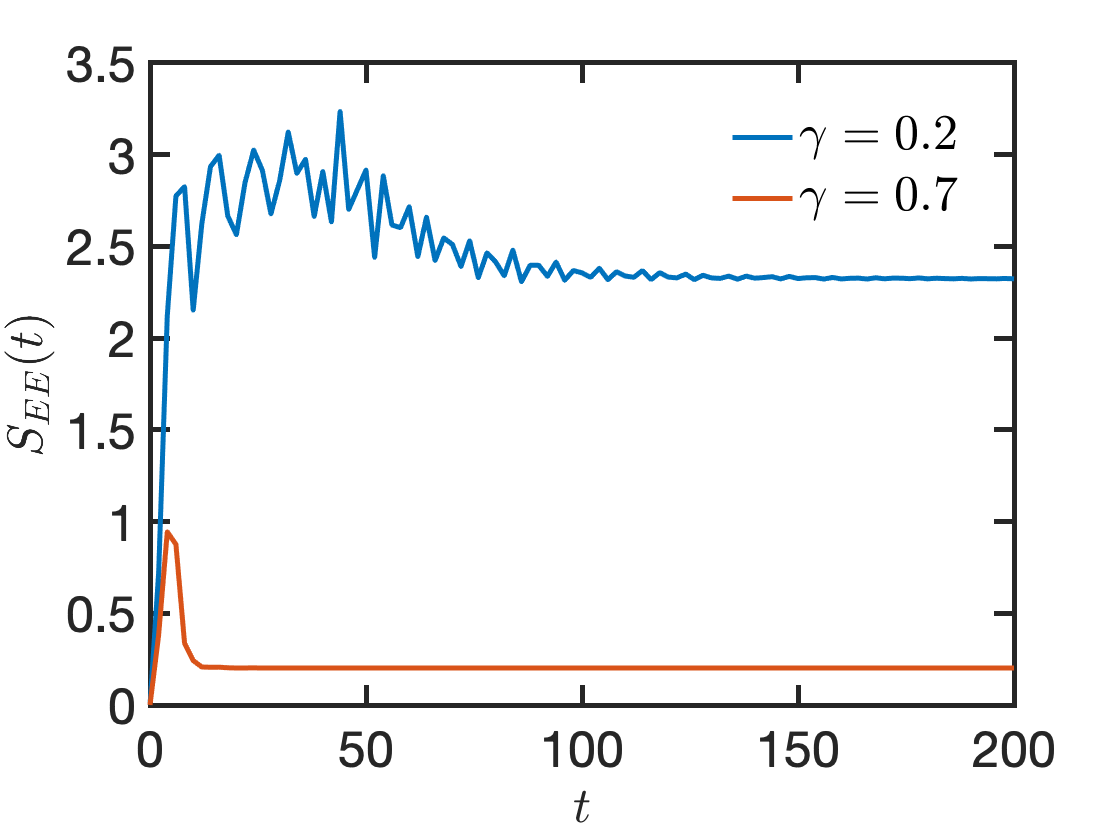}}
\subfigure[]{\includegraphics[width=0.34\textwidth]{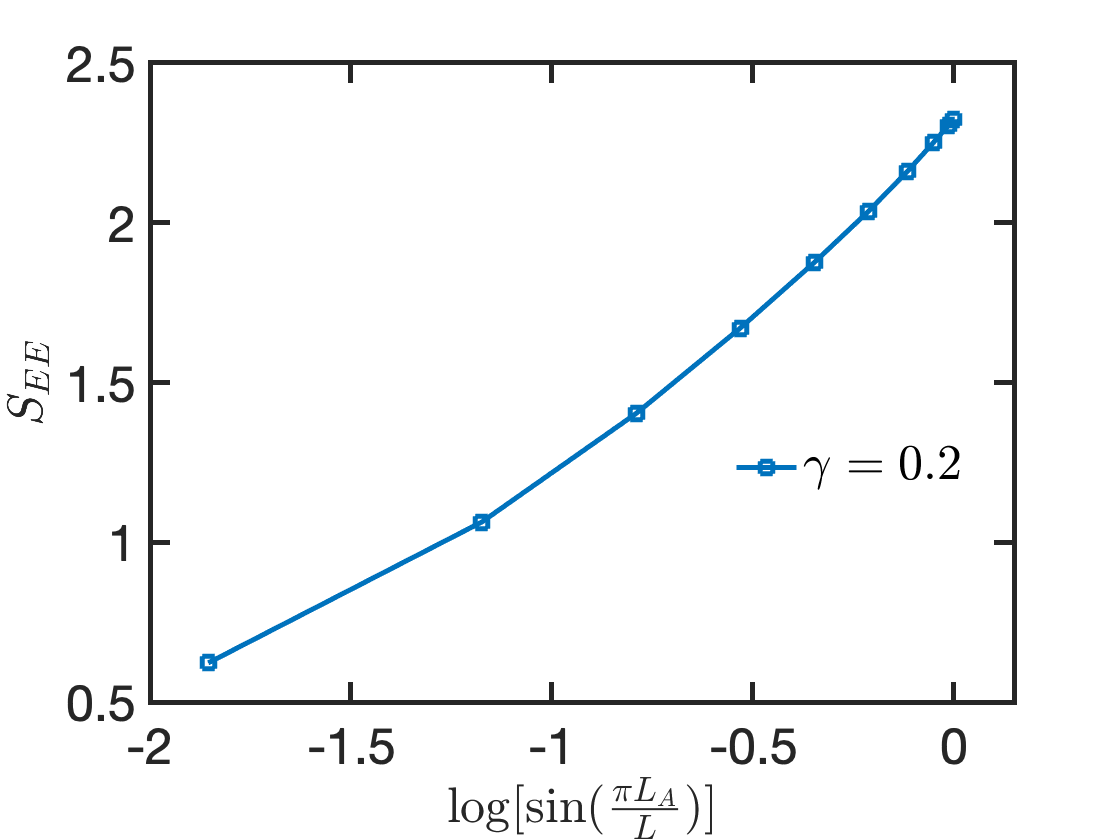}}
	\caption{(a) Time evolution of the entanglement entropy following the Floquet hybrid circuit evolution at measurement rates $\gamma < \gamma_{EE}$ and $\gamma > \gamma_{EE}$, respectively. (b) Subsystem entanglement entropy of the steady state as a function of ${\rm log}[{\rm sin}(\frac{\pi L_A}{L})]$, for $\gamma=0.2$. The system size is $L=20$ with periodic boundary condition.}
	\label{fig:circuit_log}
\end{figure}

In this section we provide additional numerical results on the forced-measurement induced entanglement transition discussed in the main text.
The circuit model consists of a unitary evolution $U = e^{-i H_0 \tau }$ with $H_0 = - \sum_{i=1}^L (J \sigma^z_i \sigma^z_{i+1} + J_2 \sigma^z_i \sigma^z_{i+2} + \Gamma \sigma^x_i)$, followed by a set of measurements given by the Kraus operators
\bea
   M_0^{(i)} = 1 - \left(1-\sqrt{1-\gamma \tau} \right) \Pi_i \quad    M_z^{(i)} = \sqrt{\gamma \tau} \Pi_i.
\eea
Since we post-select the quantum trajectory where no spin-up signal was observed, this trajectory is generated by 
\bea
    \big|\Psi_f(N\tau) \big\rangle = \frac{(M_0 U)^N | \psi_0 \rangle}{||(M_0 U)^N | \psi_0 \rangle||}, 
\eea
where $M_0 = \otimes_{i=1}^L M_0^{(i)}$ and $N \in \mathbb Z$ is the depth of the circuit, and the total evolution time is $t = N \tau$. 
We first provide additional numerical evidence supporting the volume-law scaling of the steady state entanglement for small measurement rates as shown in the main text. In Fig.~\ref{fig:circuit_log}, we plot the time evolution of the subsystem entanglement entropy following the Flouqet hybrid circuit evolution at measurement rates $\gamma < \gamma_{EE}$ and $\gamma > \gamma_{EE}$, respectively, as well as the subsystem entanglement entropy of the steady state plotted as a function of ${\rm log}[{\rm sin}(\frac{\pi L_A}{L})]$ for $\gamma=0.2$. The upward bending of the curve clearly suggests a faster-than-logarithmic growth of the bipartite entanglement, and indicates that the entanglement entropy instead obeys a volume-law scaling in this regime.

\begin{figure}
    \centering
\subfigure[]{\label{fig:area_wavefunction}
    \includegraphics[width=0.3\textwidth]{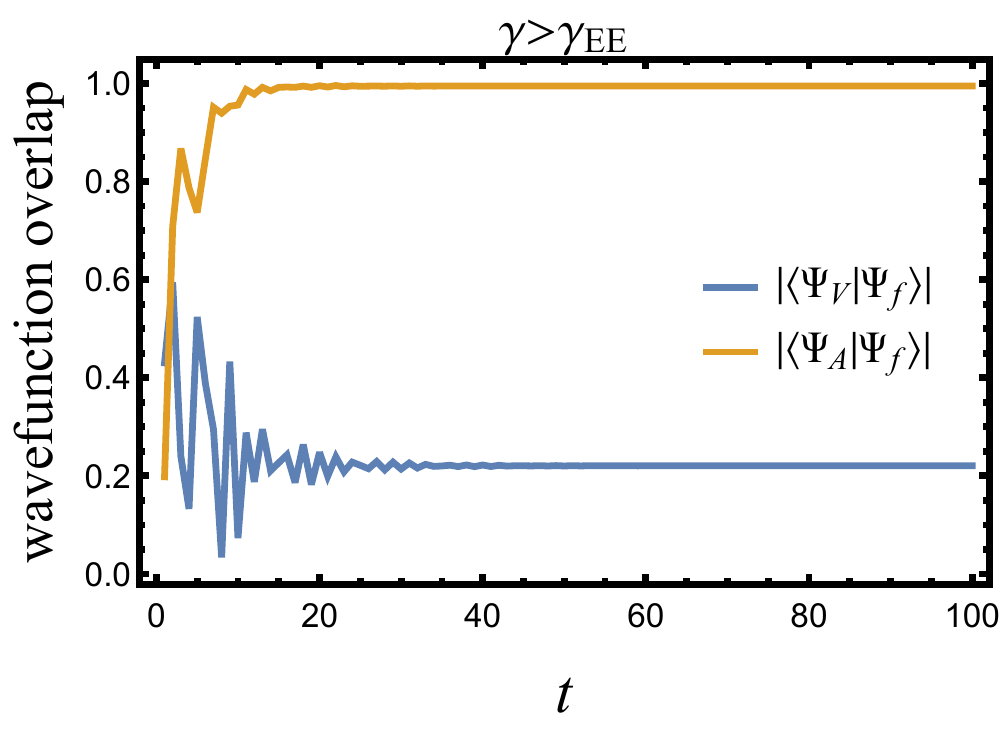}} \quad
\subfigure[]{\label{fig:volume_wavefunction}
    \includegraphics[width=0.3\textwidth]{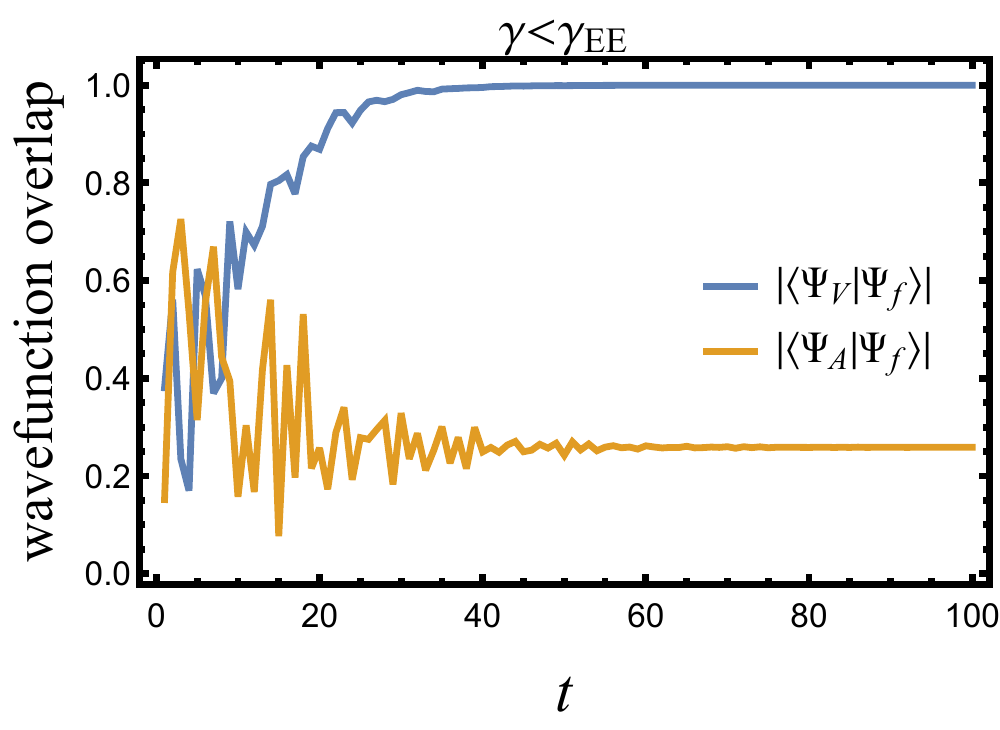}} 
    \caption{Time evolution of the wavefunction overlap in the area-law phase $\gamma > \gamma_{EE}$ (a) and volume-law phase $\gamma < \gamma_{EE}$ (b). $| \Psi_V\rangle $ ($| \Psi_A \rangle$) denotes the maximal ${\rm Im}(E)$ eigenstate of Hamiltonian~(2) in the volume-law phase (area-law phase). The parameters are $J=0.4$, $J_2=0.1$, $\Gamma=1$ and $\tau = 0.01$, and the data are obtained from a chain of size $L=10$ with periodic boundary condition.}
    \label{fig:overlap}
\end{figure}

To further substantiate the correspondence between the forced-measurement induced entanglement transition and the non-Hermitian Hamiltonian~(2), we show explicitly in Fig.~\ref{fig:overlap} the wavefunction overlaps between the final state $| \Psi_f \rangle $ and the right eigenstate of Hamiltonian~(2) with the maximal ${\rm Im}(E)$ for the area-law phase $\gamma > \gamma_{EE}$ and the volume-law phase $\gamma < \gamma_{EE}$, respectively. It is clear that the finial state from the circuit evolution has the largest overlap with the maximal ${\rm Im}(E)$ eigenstate of Hamiltonian~(2). Therefore, the long-time steady state following the forced-measurement circuit evolution is indeed controlled by the eigenstates of Hamiltonian~(2).

\section{Entanglement transition in quantum three-state Potts model}

In this section, we give an example of a non-Hermitian Hamiltonian without $\mathcal{PT}$ symmetry, and yet realizes the Yang-Lee singularity~\cite{wydro2005conformal}. We will show that a similar level crossing and entanglement transition happen in this model as well. Thus, $\mathcal{PT}$ symmetry is not a necessary condition for the mechanism studied in this work to hold.

For convenience, we write the three-state Potts model as a $\mathbb{Z}_3$ clock model:
\begin{equation}
H = - \sum_i  \left( J Z_i Z_{i+1}^\dagger + J Z_i^\dagger Z_{i+1} + X_i + X_i^\dagger  +  h \widetilde{Z}_i \right),
\label{eq:potts}
\end{equation}
where we have defined the $\mathbb{Z}_3$ clock operators satisfying 
\begin{eqnarray}
X_i^3=Z_i^3=1, \quad  X_i^\dagger = X_i^2, \quad Z_i^\dagger = Z_i^2,  \nonumber   \\
Z_i X_j = \omega^{\delta_{ij}}X_j Z_i,  \quad  Z_i X_j^\dagger = \overline{\omega}^{\delta_{ij}} X_j^\dagger Z_i,
\end{eqnarray}
with $\omega = e^{i\frac{2\pi}{3}}$ and $\overline{\omega} = \omega^2$, and $\widetilde{Z}$ is equivalent (up to a constant shift in energy) to $Z + Z^\dagger$. In the basis where $Z$ is diagonal, these clock operators take the explicit form
\begin{equation}
Z = 
\begin{pmatrix}
1 & 0 & 0 \\
0 & \omega & 0\\
0 & 0 & \bar{\omega}
\end{pmatrix},  \quad
X = 
\begin{pmatrix}
0 & 0 & 1 \\
1 & 0 & 0 \\
0 & 1 & 0
\end{pmatrix}.
\end{equation}

\begin{figure}[t]
	\centering
\subfigure[]{ 
	\includegraphics[width=0.35\textwidth]{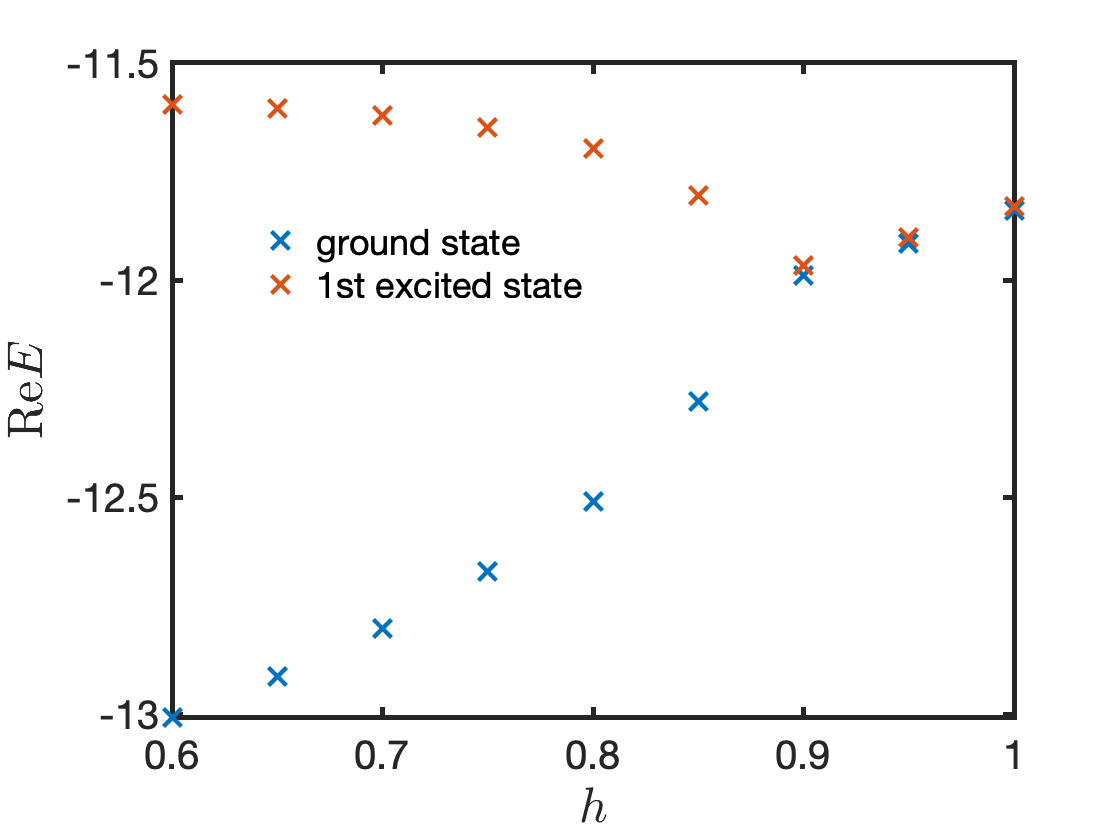}} \quad
\subfigure[]{ 
	\includegraphics[width=0.35\textwidth]{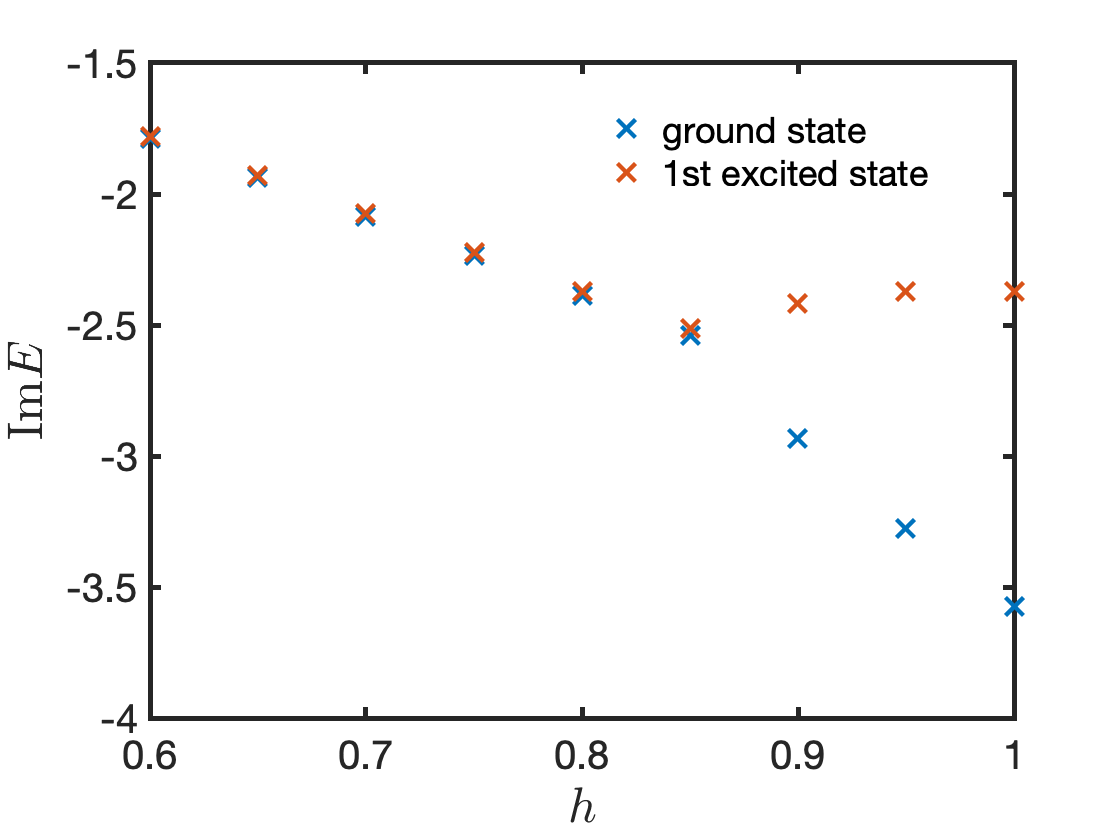}}
	\caption{The real (a) and imaginary (b) part of the eigenenergies for the ground state and first excited state of Hamiltonian~(\ref{eq:potts}). We fix ${\rm Re}(h)=0.459$ and scan vertically along the imaginary axis. The critical point where the gap closes correspond to the Yang-Lee edge singularity. We choose system size $L=6$ with periodic boundary condition.}
\label{fig:potts}
\end{figure}

\begin{figure}[t]
    \centering
    \includegraphics[width=0.35\textwidth]{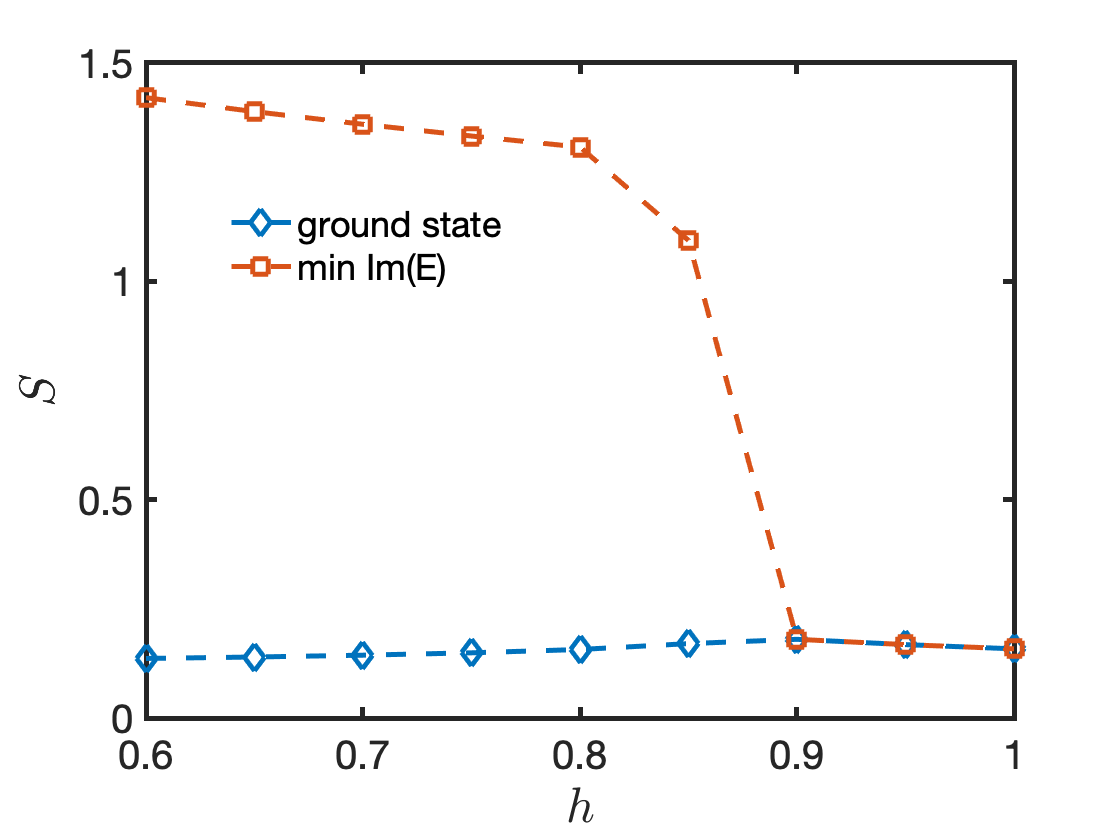}
    \caption{Entanglement entropy of the ground state and state with the smallest (largest in magnitude) ${\rm Im}(E)$. One finds a level crossing after which the ground state becomes the one with the smallest (largest in magnitude) ${\rm Im}(E)$, and the entanglement entropy jumps discontinuously. In order to observe this entanglement transition, one needs to evolve with $-H$ instead, such that the ground state (now the ceiling state) governs the long time steady state. The choice of parameters is the same as in Fig.~\ref{fig:potts}.}
    \label{fig:potts_entropy}
\end{figure}

In Fig.~\ref{fig:potts}, we plot the real and imaginary part of the eigenenergies for the ground state and first excited state of Hamiltonian~(\ref{eq:potts}) as $h$ is varied. We find that the gap between these two states closes at a critical $h$ corresponding to the Yang-Lee edge singularity, and then reopens such that the eigenvalues split along the imaginary axis. Since the ground state enters a $\mathbb{Z}_3$ symmetry broken phase with a large magnetization $\langle \widetilde{Z} \rangle$, which is proportional to the imaginary part of the eigenenergy, we expect that the ground state will dominate the long-time steady state as $h$ increases, and a level crossing must happen in this process. In Fig.~\ref{fig:potts_entropy}, we plot the entanglement entropy of the ground state as well as eigenstate with the smallest (largest in magnitude) ${\rm Im}(E)$. The result indeed confirms the existence of a level crossing after which the scaling of the steady state entanglement changes discontinuously. Notice that in order to observe this entanglement transition, one needs to evolve with $-H$ instead, such that the ground state (now the ceiling state) will have the {\it largest} imaginary eigenenergy, and dominate the long-time steady state.

\section{Non-Hermitian Hamiltonian without Yang-Lee edge singularity}

We show in the main text that the Yang-Lee edge singularity triggers a level crossing leading to an entanglement transition. The crucial ingredient of this mechanism is the gap closing of the ground state followed by a $\mathcal{PT}$ symmetry breaking, when the ground state energy becomes complex. As the non-Hermitian part is increased further, the ground state finally takes over and dominates in the long-time steady state. We refer to this as the ground state $\mathcal{PT}$ transition. In this section, we contrast this scenario with situations where there is no ground state $\mathcal{PT}$ transition. 

\begin{figure}
	\centering
\subfigure[]{ \label{fig:imag1}
	\includegraphics[width=0.36\textwidth]{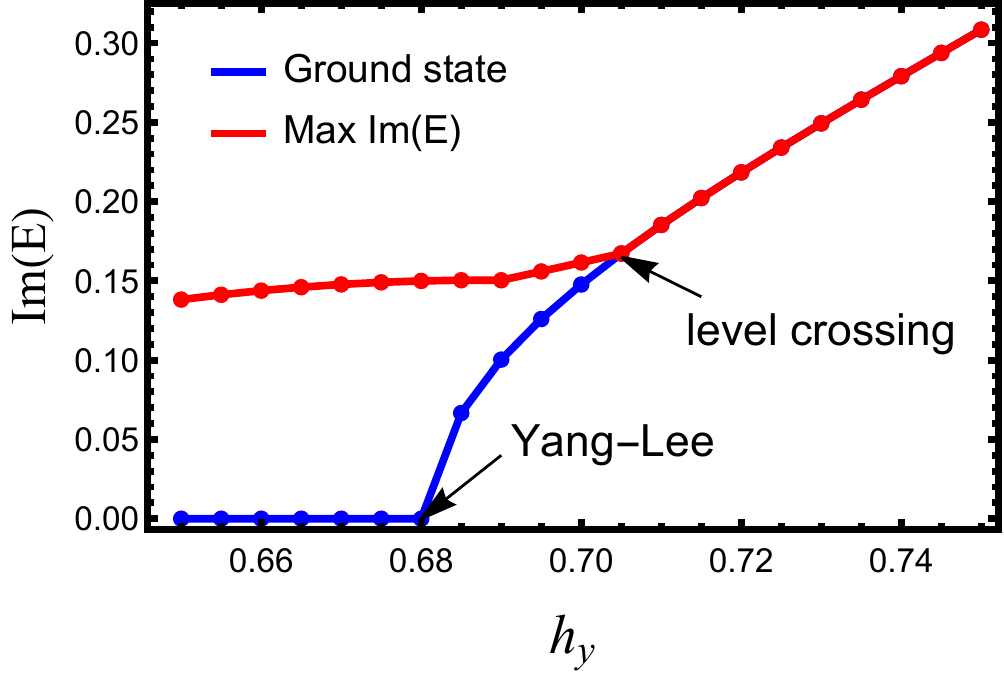}} \quad
\subfigure[]{ \label{fig:order1}
	\includegraphics[width=0.35\textwidth]{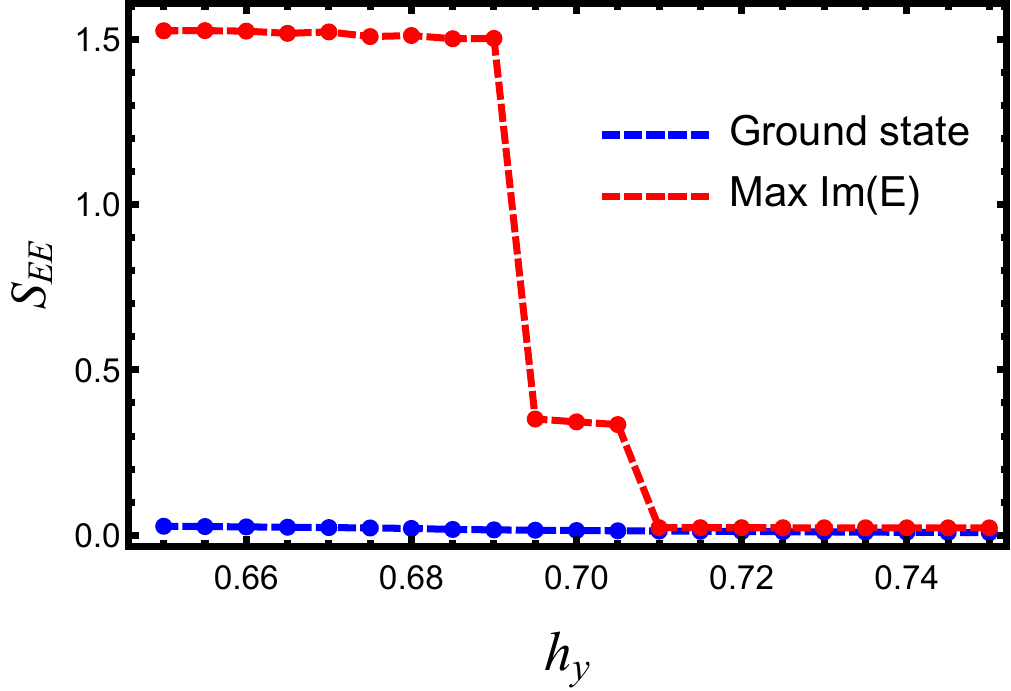}}
	\caption{The ${\rm Im}(E)$ (a) and half-chain entanglement entropy (b) of the ground state and the state with maximal ${\rm Im}(E)$. The parameters are $J = 0.4, J_2=0, \Gamma = 1, h_z = 0.1 $, and the data are obtained from exact diagonalization of a chain of length $L=10$ with periodic boundary condition.}
\end{figure}

Interestingly, both scenarios can be realized in Hamiltonian~(2) by taking the longitudinal field $h_z$ to be either real or imaginary. We first show the entanglement transition in Hamiltonian~(2) following a different cut in the phase diagram, by keeping $h_z$ fixed and varying $h_y$. In Fig.~\ref{fig:imag1} and Fig.~\ref{fig:order1}, we plot the ${\rm Im}(E)$ and the half-chain entanglement entropy for both the ground state and the state with maximal ${\rm Im}(E)$, respectively. They are similar to Fig.~2(c) and Fig.~2(d) in the main text.

To explore the other scenario, it is most convenient to make the longitudinal magnetic field real $ i h_z \rightarrow h_z'$, $h_z' \in \mathbb R$. Namely, we consider (setting $J_2=0$)
\bea\label{eq:hamiltonian2}
	H = - \sum_{i=1}^L (J \sigma^z_i \sigma^z_{i+1} + \Gamma \sigma^x_i + h_z' \sigma^z + i h_y \sigma^y ), \quad h_z' \in \mathbb R.
\eea 
This is precisely the model recently studied in Ref.~\cite{gopalakrishnan2020entanglement}.
The Hamiltonian explicitly breaks the $\mathcal{PT}$ symmetry associated with Hamiltonian~(2). Nevertheless, it does preserve the $\mathcal T$ symmetry, because the Hamiltonian is purely real. Its eigenvalues are either real or pairs that are related by complex conjugation. 
\begin{figure}
	\centering
\subfigure[]{ \label{fig:imag2}
	\includegraphics[width=0.35\textwidth]{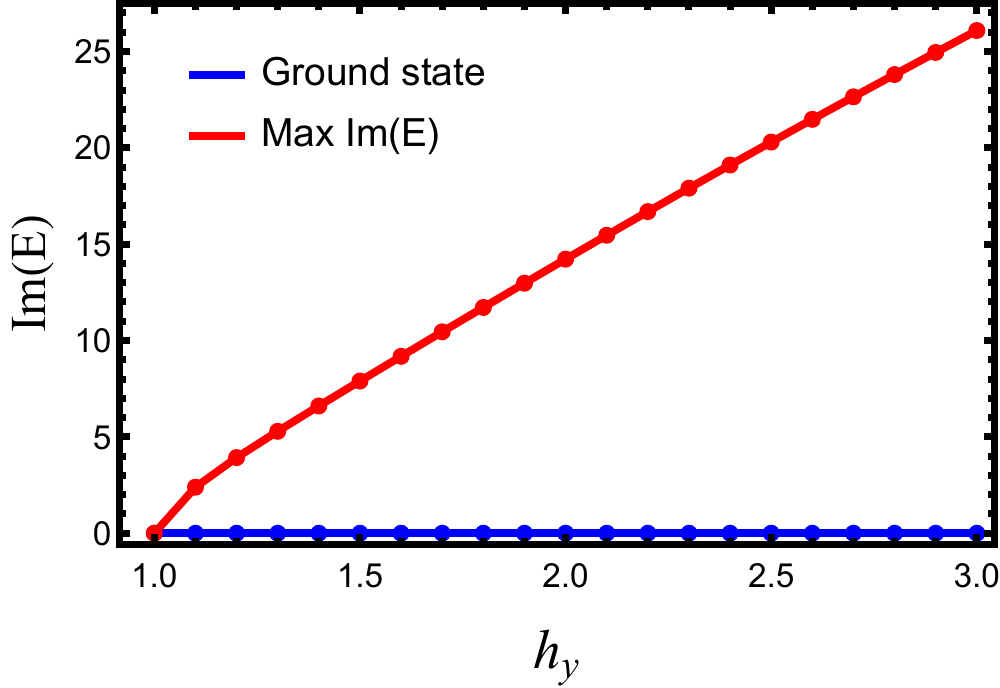}} \quad
\subfigure[]{ \label{fig:order2}
	\includegraphics[width=0.36\textwidth]{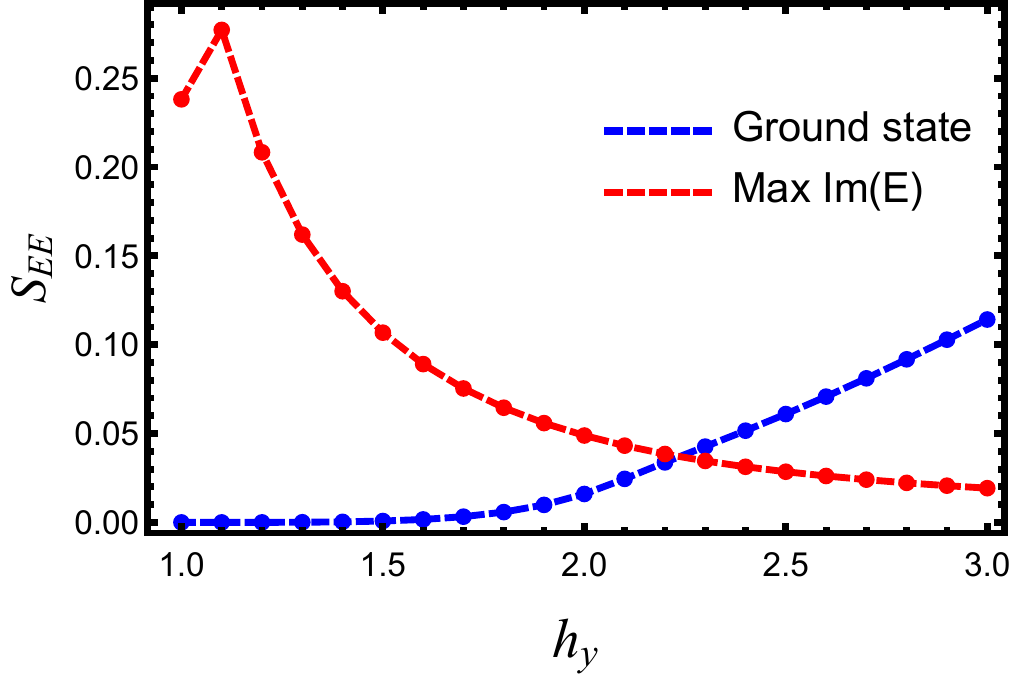}}
	\caption{The ${\rm Im}(E)$ (a) and half-chain entanglement entropy (b) of the ground state and the state with maximal ${\rm Im}(E)$. The parameters are $J = 0.4, J_2=0, \Gamma = 1, h_z' = 1.1 $, and the data are obtained from exact diagonalization of a chain of length $L=10$ with periodic boundary condition.}
\end{figure}

In Fig.~\ref{fig:imag2}, we plot the ${\rm Im}(E)$ for both the ground state and the maximal imaginary energy eigenstate. Clearly, the ground state energy in this case is always real. This is expected because in the absence of a non-unitary critical point, the ground state gap does not close in general. Therefore, the ground state energy must be real due to the $\mathcal T$ symmetry. The half-chain entanglement entropy of the maximal ${\rm Im}(E)$ eigenstate, as shown in Fig.~\ref{fig:order2}, decreases continuously as $h_y$ varies, once the spectrum becomes complex.

\end{document}